\documentclass[12pt]{article}

\usepackage{authblk}

\usepackage[super,sort&compress]{natbib}
\setcitestyle{super,comma}
\usepackage{bibunits} \defaultbibliographystyle{naturemag}
\defaultbibliography{Biblio}

\usepackage[mathlines]{lineno}
\usepackage{upgreek} 
\usepackage{tabularx}
\usepackage{times}
\usepackage[utf8]{inputenc}
\usepackage[T1]{fontenc}
\usepackage{caption}
\usepackage{sectsty}
\sectionfont{\fontsize{13}{15}\selectfont}
\subsectionfont{\fontsize{12}{15}\selectfont}
\usepackage{nameref}
\usepackage{titlesec}

\usepackage{amsmath}
\usepackage{amssymb}
\usepackage{graphicx}
\usepackage{color}
\usepackage{bm}
\usepackage{setspace}

\usepackage[usenames,dvipsnames]{xcolor}

\usepackage{abstract}

\makeatletter
\renewcommand{\maketitle}{\bgroup\setlength{\parindent}{0pt}
\begin{flushleft}
  \textbf{\@title}\\[2em]
  \@author
\end{flushleft}\egroup
}
\makeatother

\title{\noindent {\bf \large Visualizing impurity-driven scattering phase textures in EuCd$_2$As$_2$}}

\author[1]{\normalsize Raquel Sánchez-Barquilla}
\author[2]{\normalsize Rafael Pineda Medina}
\author[1]{\normalsize Pablo Garc\'ia Talavera}
\author[3]{\normalsize Adrian Valadkhani}
\author[1]{\normalsize Edwin Herrera}
\author[4]{\normalsize Brinda Kuthanazhi}
\author[4]{\normalsize Lin-Lin Wang}
\author[4]{\normalsize Sergey L. Bud'ko}
\author[4]{\normalsize Paul C. Canfield}
\author[3]{\normalsize Roser Valentí}
\author[1]{\normalsize Isabel Guillam\'on}
\author[2]{\normalsize William J. Herrera}
\author[5]{\normalsize Alfredo Levy Yeyati}
\author[1]{\normalsize Hermann Suderow}

\affil[1]{\small \it Laboratorio de Bajas Temperaturas y Altos Campos Magn\'eticos, Departamento de F\'isica de la Materia Condensada and Instituto Nicol\'as Cabrera, Universidad Aut\'onoma de Madrid, E-28049 Madrid, Spain.}
\affil[2]{\small \it Departamento de F\'isica, Universidad Nacional de Colombia, 111321 Bogot\'a, Colombia.}
\affil[3]{\small \it Institut für Theoretische Physik, Goethe Universität Frankfurt, Max-von-Laue Strasse 1, 60438 Frankfurt am Main, Germany.}
\affil[4]{\small \it Ames Laboratory and Department of Physics \& Astronomy, Iowa State University, Ames, IA 50011.}
\affil[5]{\small \it Departamento de F\'isica Te\'orica de la Materia Condensada, Instituto Nicol\'as Cabrera and Condensed matter Physics Center (IFIMAC), Universidad Aut\'onoma de Madrid, E-18049, Madrid, Spain}

\begin{document}

\baselineskip24pt


\maketitle 	
\noindent{\bf 
Understanding how disorder modifies electronic states in magnetic semiconductors is important for controlling spin-dependent transport and topological responses. Here we use scanning tunneling microscopy to visualize scattering phase textures around Eu interstitials in EuCd$_2$As$_2$. By isolating a single surface wavevector we reconstruct spatial phase maps of the local density of states and identify phase dislocations characterized by $2\pi$ winding around impurity sites. These phase singularities emerge systematically within charge puddles generated by Eu interstitials and their positions evolve with bias voltage. We show that their spatial structure is consistent with interference between multiple scattering channels, including contributions from spin–orbit coupling. We provide a model which reproduces phase dislocations and relates the decay of the phase gradient to the relative strength of spin–orbit and scalar scattering. Our results establish a route to access the phase of electronic scattering in real space and study the role of local disorder and spin–orbit interactions in shaping electronic states in quantum materials.}


\paragraph*{Scattering in magnetic semiconductors.}

Electronic scattering processes govern charge transport, spin dynamics and the emergence of collective quantum phenomena in condensed matter systems. In the presence of spin--orbit coupling (SOC), scattering acquires an internal phase structure that reflects the entanglement of spin and momentum degrees of freedom, underpinning effects such as weak antilocalization \cite{Hikami1980}, the spin Hall effect \cite{Sinova2015} and topological transport \cite{Hasan2010,Qi2011}. Although these phenomena ultimately originate from phase accumulation in electronic wavefunctions, experimental access to the scattering phase in real space remains severely limited.

Here we analyze textures in the electronic structure of the magnetic semiconductor EuCd$_2$As$_2$ around Eu interstitials. EuCd$_2$As$_2$ is particularly attractive because its electronic structure is extremely sensitive to Eu content and local disorder\cite{PhysRevLett.131.186704,PhysRevMaterials.7.034402,shi2023absence,nishihaya2024intrinsic,nahyun2020,gati2021, yu2023}. As we show below, Eu interstitials can strongly modify the low-energy electronic states, producing localized charge puddles and spatially varying potentials. Understanding how these defects scatter electrons is essential for controlling transport and possible topological responses in magnetic semiconductors.

Scanning tunneling microscopy (STM) and quasiparticle interference (QPI) techniques have provided detailed insight into the dispersion and symmetry of electronic states by probing spatial modulations of the local density of states (LDOS) \cite{Pascual2004,Hoffman2002}. However, these approaches predominantly reveal the amplitude of electronic wavefunctions, while the associated phase information is typically averaged out or inferred indirectly through symmetry considerations. Accessing the spatial structure of the scattering phase would provide a powerful new probe of quantum materials, enabling direct visualization of phase winding, Berry-phase effects and spin-dependent scattering processes at the nanoscale \cite{RevModPhys.81.1495}.

\paragraph*{Scattering phase textures.}

Recent work has shown that phase information can be encoded in real-space LDOS modulations when a well-defined scattering wavevector is present, allowing reconstruction of an effective phase field from spatially resolved measurements  \cite{Lawler2010}. In this geometric phase analysis framework \cite{10.1093/jmicro/dfaa063,10.1017/S1431927622000125,HYTCH1998131,Lawler2010,Herrera2021}, the LDOS is modeled locally as an amplitude- and phase-modulated cosine, $\rho(\mathbf{r}) \approx \rho_0(\mathbf{r}) + A(\mathbf{r}) \cos(\mathbf{q}_0\cdot\mathbf{r}+\phi(\mathbf{r}))$, where $A(\mathbf{r})$ and $\phi(\mathbf{r})$ vary slowly compared to the underlying periodicity $2\pi/|\mathbf{q}_0|$. Extracting $\phi(\mathbf{r})$ then amounts to demodulating this signal: shifting the Fourier transform of $\rho(\mathbf{r})$, $\rho_F(\mathbf{q})$, by $\mathbf{q}_0$ and retaining only the low-q content around the shifted origin isolates a single sideband, whose inverse transform is the complex analytic field $(A(\mathbf{r})/2)e^{i\phi(\mathbf{r})}$. This filtering step breaks the hermiticity $\rho_F(\mathbf{q})=-\rho_F^*(\mathbf{q})$ of the Fourier transform of the original real-valued signal, since the conjugate sideband near $-\mathbf{q}_0$ is discarded, leaving a genuinely complex field from which an unambiguous phase can be read off. Points where $A(\mathbf{r})\to 0$ are phase singularities: there the modulation locally vanishes, and $\phi(\mathbf{r})$ is undefined and winds by integer multiples of $2\pi$ around the point singularity. These phase dislocations provide a direct real-space signature of the interference between different scattering channels. In systems with SOC, scattering processes can generate phase structures due to the momentum-dependent coupling between spin and orbital degrees of freedom \cite{Bercioux2015,Winkler2003}, suggesting that phase-sensitive measurements could reveal new aspects of spin-dependent electronic transport.

\begin{figure}
	\includegraphics[width=1\columnwidth]{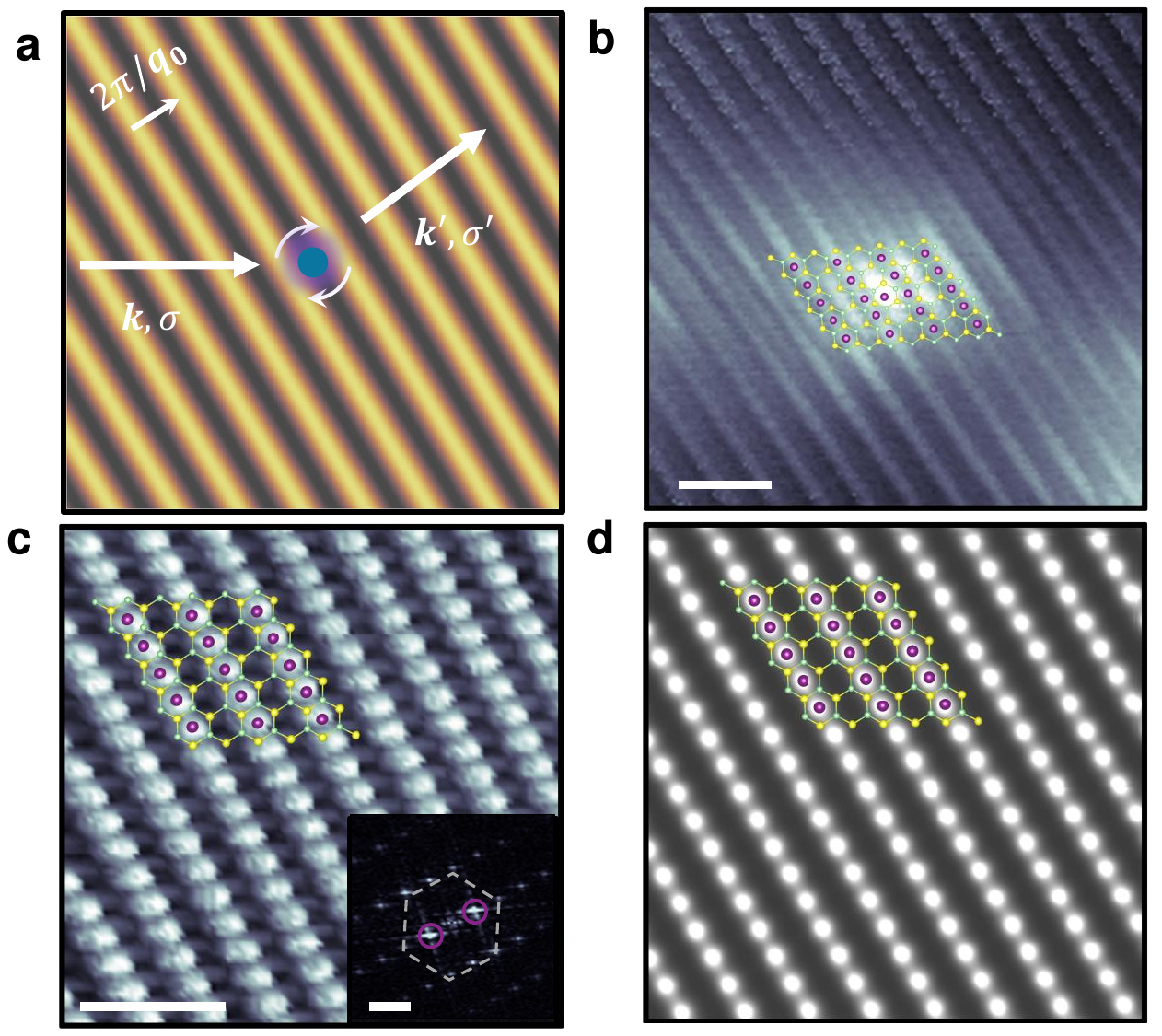}
	\caption{~{\bf{One-dimensional Eu rows on the surface of EuCd$_2$As$_2$ with and without Eu interstitial impurities.}} {\bf a.}~Schematic representation of a one-dimensional pattern of fringes, with wavevector $q_0$. We also represent schematically spin orbit scattering by an impurity (blue disk). An electronic wavefunction with wavevector $\mathbf{k}$ and spin $\sigma$ scatters at the impurity site. {\bf b.}~Experimentally obtained STM topographic image showing a Eu interstitial. The horizontal white line is 3~nm long and the grey scale corresponds to equivalent changes in height of approximately 0.1~nm. {\bf c} STM topographic image of EuCd$_2$As$_2$ in a field of view free of interstitials. The horizontal white line is 1~nm long and the grey scale corresponds to equivalent changes in height of approximately 0.1~nm. In the inset we show a FFT map, where the horizontal white line is 3~nm$^{-1}$ long. We highlight with violet circles the Bragg peaks from the Eu rows seen in the real space image. The grey hexagon provides the orientation and size of the underlying Cd sublattice. {\bf d}~STM topography calculated by DFT. We provide in all images the atomic positions of EuCd$_2$As$_2$ projected to the cleaving plane as colored disks (Eu in violet, As in green and Cd in yellow). The maps {\bf b,c} have been taken at a bias voltage of $+$100~mV and 0.4~nA current.}
	\label{FigureTopo}
\end{figure}

Here, using low-temperature STM, we map the LDOS around Eu interstitials in EuCd$_2$As$_2$. To access phase information, rather than focusing on energy–momentum dispersion as in conventional quasiparticle interference (QPI), we analyze spatial modulations of the LDOS, $\rho(\mathbf{r})$, characterized by a dominant wavevector, $q_0$ (Fig.~\ref{FigureTopo}{\bf a}). Thanks to having a well defined wavevector $\mathbf{q}_0$, we can filter out other components as described above and obtain the phase map $\phi(\mathbf{r})$. When the phase is uniform, $\phi(\mathbf{r}) = \phi_0$, the resulting LDOS pattern consists of parallel fringes with wavelength $2\pi/|\mathbf{q}_0|$ (Fig.~\ref{FigureTopo}{\bf a}). As we show below, spin orbit scattering at an impurity, schematically represented in Fig.~\ref{FigureTopo}{\bf a} by arrows and the blue point, leads to a scattering pattern phase with phase windings of $\pm 2\pi$ located near the impurity.

\paragraph*{Topography and surface of EuCd$_2$As$_2$.} EuCd$_2$As$_2$ is a layered compound in which Eu layers enclose As–Cd–Cd–As sheets. The Eu magnetic moments order ferromagnetically within each plane and antiferromagnetically between adjacent planes below $T_N = 9$ K. We focus here on atomically flat Eu-terminated surfaces, which exhibit a reconstructed structure of quasi-one-dimensional Eu rows. Additional details on the crystal structure and alternative surface terminations (Extended Data Figure\,\ref{SuplLattice}) are provided in Methods Sections A and B.

Figure~\ref{FigureTopo}{\bf b} shows an STM topography (see Methods Section C for more details) of such a surface containing a single impurity, identified by the accumulation of charge (bright contrast). We overlay schematically the Eu rows, the interstitial Eu atom and the underlying hexagonal Cd and As sublattices. The row-like pattern in the vicinity of the impurity appears distorted. This reflects the fact that the topographic image corresponds to the energy-integrated density of states up to the applied bias voltage. As demonstrated below, the LDOS at fixed energy can be described as $\rho(E,\mathbf{r}) \propto \cos(\mathbf{q}_0 \cdot \mathbf{r} + \phi(E,\mathbf{r}))$, where $\mathbf{q}_0$ is set by the Eu row periodicity and the phase $\phi(E,\mathbf{r})$ varies spatially. In particular, we observe phase slips, corresponding to $2\pi$ windings of $\phi(E,\mathbf{r})$, whose positions evolve strongly with energy. For reference, we provide a STM topography experiment in Fig.\,\ref{FigureTopo}{\bf c} and a density functional theory (DFT) calculation of the DOS on relaxed slabs close to the Fermi energy in Fig.\,\ref{FigureTopo}{\bf d} in a field of view without Eu interstitials (see Methods Section D for details). We observe a well-defined pattern of one-dimensional rows, following the unperturbed surface Eu lattice of EuCd$_2$As$_2$.

\paragraph*{Bandstructure of EuCd$_2$As$_2$ with Eu interstitials.}

\begin{figure*}
	\includegraphics[width=1\textwidth]{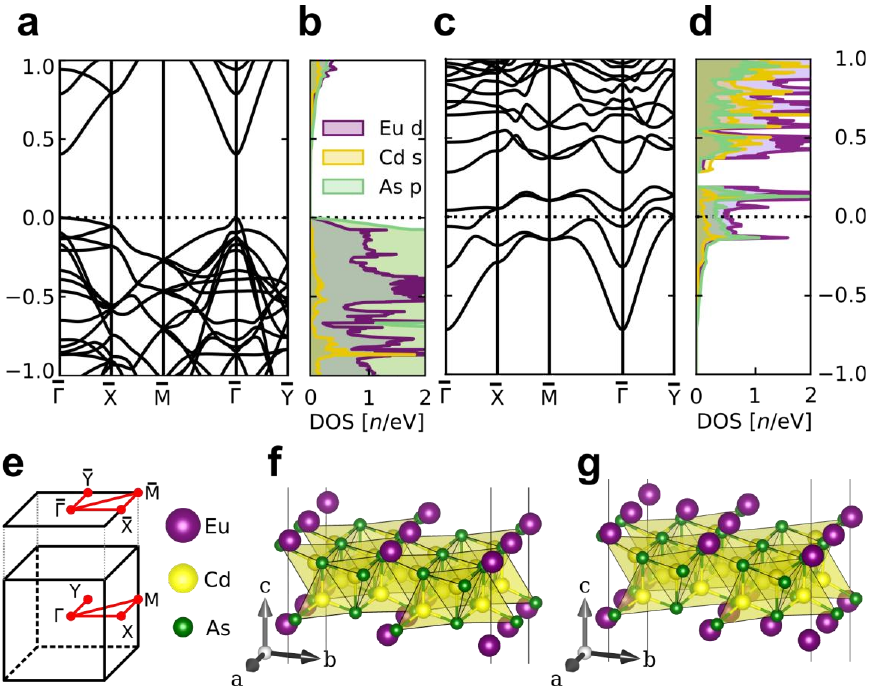}
	\caption{~{\bf Electronic bandstructure of EuCd$_2$As$_2$.} {\bf a} DFT calculations of the band structure of a surface terminated by Eu rows, similar to the one observed in the experiment (see also Fig.\,\ref{FigureTopo}{\bf c,d}). {\bf b} DOS corresponding to {\bf a}. Colors indicate the contributions from different orbitals, green for As~4p (dominant contribution around the band gap), violet for Eu~5d and yellow for Cd~5s. {\bf c} Band structure with an Eu interstitial (see also Fig.\,\ref{FigureTopo}{\bf b}). Note the significant shift downward, as compared to {\bf a}. {\bf d} DOS corresponding to {\bf c}. {\bf e} Brillouin Zone of EuCd$_2$As$_2$ showing the main high symmetry directions of the a-b plane ($\Gamma$, $X$, $Y$, $M$) and their projection to the cleaved surface ($\overline{\Gamma}$, $\overline{X}$, $\overline{Y}$, $\overline{M}$). {\bf f} Slab of a surface of EuCd$_2$As$_2$ terminated with Eu rows. {\bf g} Slab with Eu rows termination including a Eu interstitial in between rows. We use the same color scheme for the atoms in {\bf e,f,g} and the partial DOS in {\bf b,d}.}
	\label{Figure2}
\end{figure*}

Additional  DFT calculations of the electronic structure of EuCd$_2$As$_2$ are presented in Fig.~\ref{Figure2}. The surface band structure, obtained from slabs terminated by Eu rows (Fig.~\ref{Figure2}{\bf a,b}), reveals a sizable gap at the $\Gamma$ point, with no states up to approximately 0.4 eV. The valence band is predominantly of As~4p character, with a significant contribution from Eu~5d states. Consistent with these calculations, tunneling conductance curves measured far from impurities (see Supplementary Information Sections I and II) suggest a strongly suppressed density of states for empty states up to several tens of meV, in agreement with the semiconducting gap predicted by DFT.

Remarkably, a single Eu interstitial dramatically reconstructs the local electronic structure, shifting Eu-derived bands by approximately 1.5~eV and transforming the gapped surface into a low-energy metallic state (Fig.\,\ref{Figure2}{\bf c},{\bf d}). The electronic character of the bands crossing the Fermi level is now predominantly Eu~5d. The bands have a pronounced one-dimensional character, with features along $\Gamma-X$. Depending on the position of the Fermi level, we identify one central electron pocket at $\Gamma$ and pockets at $X$, although their size is strongly dependent on the position of the Fermi level.

\begin{figure*}
	\includegraphics[width=1\textwidth]{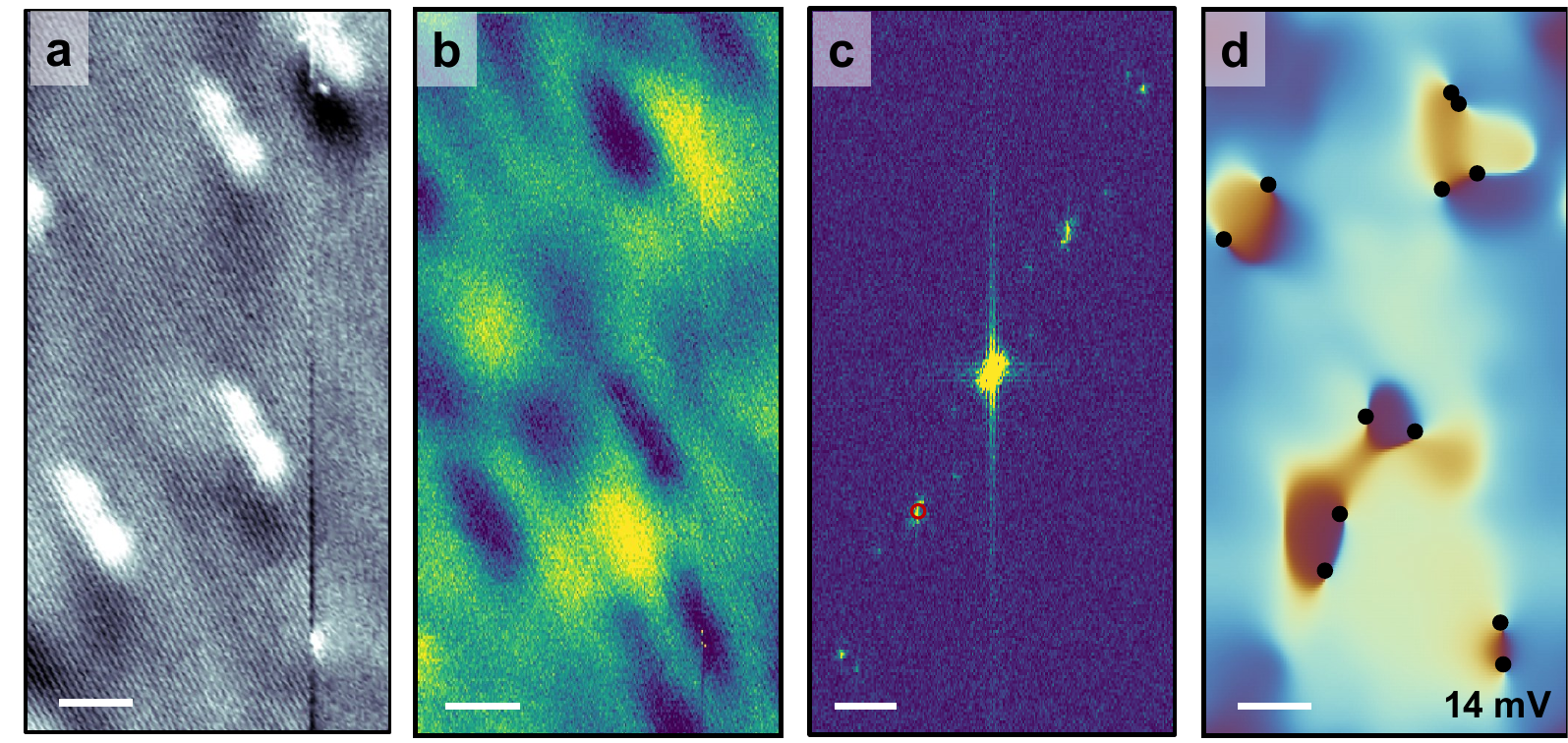}
	\caption{~{\bf Patterns of the DOS and observation of scattering textures.} {\bf a} Topographic map taken at a bias voltage of $+$100~mV and a tunneling current of 0.2~nA. We identify the one-dimensional row-like structure of the Eu $2\times1$ reconstruction (see Figs.\,\ref{FigureTopo},\ref{Figure2}) and the extended charge puddles around Eu impurities (areas in white). {\bf b} Tunneling conductance map at zero bias. We discuss the full tunneling conductance as a function of the bias voltage in Supplementary Information Section II. {\bf c} Fourier transform of {\bf b}. We mark with a red circle one of the Bragg peaks due to the Eu rows, at ${\mathbf Q}_{1\mathrm{D}}$. {\bf d} Local phase $\phi$ of the modulation of the tunneling conductance at ${\mathbf Q}_{1\mathrm{D}}$ for 14~mV. The phase winds by $2\pi$ on a circle around the black points. The full bias voltage dependence of the phase is provided in Extended Data Figures~\ref{SuplPhaseFullMaps0T},~\ref{SuplPhaseFullMaps14T} and  the full bias dependence of the tunneling conductance in the Supplementary Information Section II. Horizontal white scale bar is 10~nm long.}
	\label{FigurePhaseMap}
\end{figure*}

\paragraph*{Charge puddles and phase textures at impurities in EuCd$_2$As$_2$.} The most commonly observed situation in STM measurements on EuCd$_2$As$_2$ surfaces is illustrated in Fig.~\ref{FigurePhaseMap}{\bf a}, where a high density of closely spaced impurities gives rise to extended charge puddles (see Extended Data Figure\,\ref{SuplExampleFilter}, \ref{MapsNoImp}, \ref{SuplPhaseFullMaps0T}, \ref{SuplPhaseFullMaps14T} for other examples). The corresponding tunneling conductance maps, for instance at zero bias (Fig.~\ref{FigurePhaseMap}{\bf b}), exhibit a  pattern of peaks and troughs whose spatial distribution and characteristic length scales evolve with bias voltage (full maps are provided in the Supplementary Information Sections I and II). Spectroscopy at fixed positions near impurities reveals multiple in-gap resonances, whose energies and amplitudes vary strongly across the sample. Similar spatially inhomogeneous impurity states have been reported in doped semiconductors\cite{MORGENSTERN20121795,PhysRevB.80.161311,doi:10.1126/science.aag1715,10.1063/1.4942517,PhysRevB.92.085140,Morgenstern2012}.

However, in contrast to those previous studies, we identify here a robust periodic modulation in the tunneling conductance at all bias voltages, with a well-defined wavevector corresponding to the Eu rows, ${\mathbf Q}_{1\mathrm{D}}$. The Fourier transform of the zero-bias conductance map (Fig.~\ref{FigurePhaseMap}{\bf c}) exhibits sharp Bragg peaks at ${\mathbf Q}_{1\mathrm{D}}$, indicating a dominant one-dimensional modulation. This allows us to express the tunneling conductance as $\sigma(V,\mathbf{r}) = A_{\mathrm{mod}}(\mathbf{r},V)\cos\big({\mathbf Q}_{1\mathrm{D}}\cdot\mathbf{r} + \phi(\mathbf{r},V)\big)$, where $\mathbf{r}=(x,y)$ denotes the surface position, and both the amplitude $A_{\mathrm{mod}}$ and phase $\phi$ depend on position and bias voltage. By applying a Fourier filter around ${\mathbf Q}_{1\mathrm{D}}$, highlighted by the red circle in Fig.~\ref{FigurePhaseMap}{\bf c}, we isolate this contribution and reconstruct the spatial phase $\phi(\mathbf{r},V)$ (Fig.~\ref{FigurePhaseMap}{\bf d}, see Methods Section E). The resulting maps reveal numerous phase slip pairs distributed across the field of view, predominantly located near impurity sites. The correlation between phase-slip locations and Eu interstitials suggests that phase mapping can be used as a nanoscale diagnostic of defect-generated spin-orbit scattering. This opens possibilities for characterizing and engineering disorder landscapes in magnetic semiconductors.

The emergence of a one-dimensional modulation at ${\mathbf Q}_{1\mathrm{D}}$ is expected, as the Eu rows define a well-characterized scattering potential on the surface. However, the observation of phase slips in the tunneling conductance departs from the behavior of conventional scalar scattering processes. Ordinary scalar scattering produces a standing-wave pattern with a well-defined global phase. In this case the LDOS modulation can vanish along nodal lines, but its phase remains single-valued. Spin-orbit scattering introduces an additional component with a different angular dependence. The interference of the scalar and spin-orbit channels leads to a contribution to the LDOS of the form $\mathcal{A}(\mathbf{r})\cos\big(Q_{1\mathrm{D}}x\big)+\mathcal{B}(\mathbf{r})\cos\big(Q_{1\mathrm{D}}x\big) \sin\theta$, which can be written as $Re[\Phi(r,\theta)e^{iQ_{1\mathrm{D}}x}]$, with $\Phi(r,\theta)=\mathcal{A}(\mathbf{r})-i\mathcal{B}(\mathbf{r})\sin\theta = \vert \Phi(\mathbf{r},\theta) \vert e^{i\phi(\mathbf{r},\theta)}$ and gives a dislocation for $\mathcal{A}(\mathbf{r})=0$ and $\sin\theta=0$ in which $\phi$ is undefined and winds by $2\pi$ (details in Methods Section F). Non-trivial phase accumulation due to Berry-phase singularities (for example, Dirac cones) also provides dislocations in the LDOS\cite{CRPHYS202122S41330,scienceaad8038}. However, the semiconducting band structure of EuCd$_2$As$_2$ shows no evidence of such topological features (see Supplementary Information Section III and Fig.~\ref{Figure2}), leaving spin-orbit scattering as the most likely mechanism for the observed phase winding.

\begin{figure*}
	\includegraphics[width=1\textwidth]{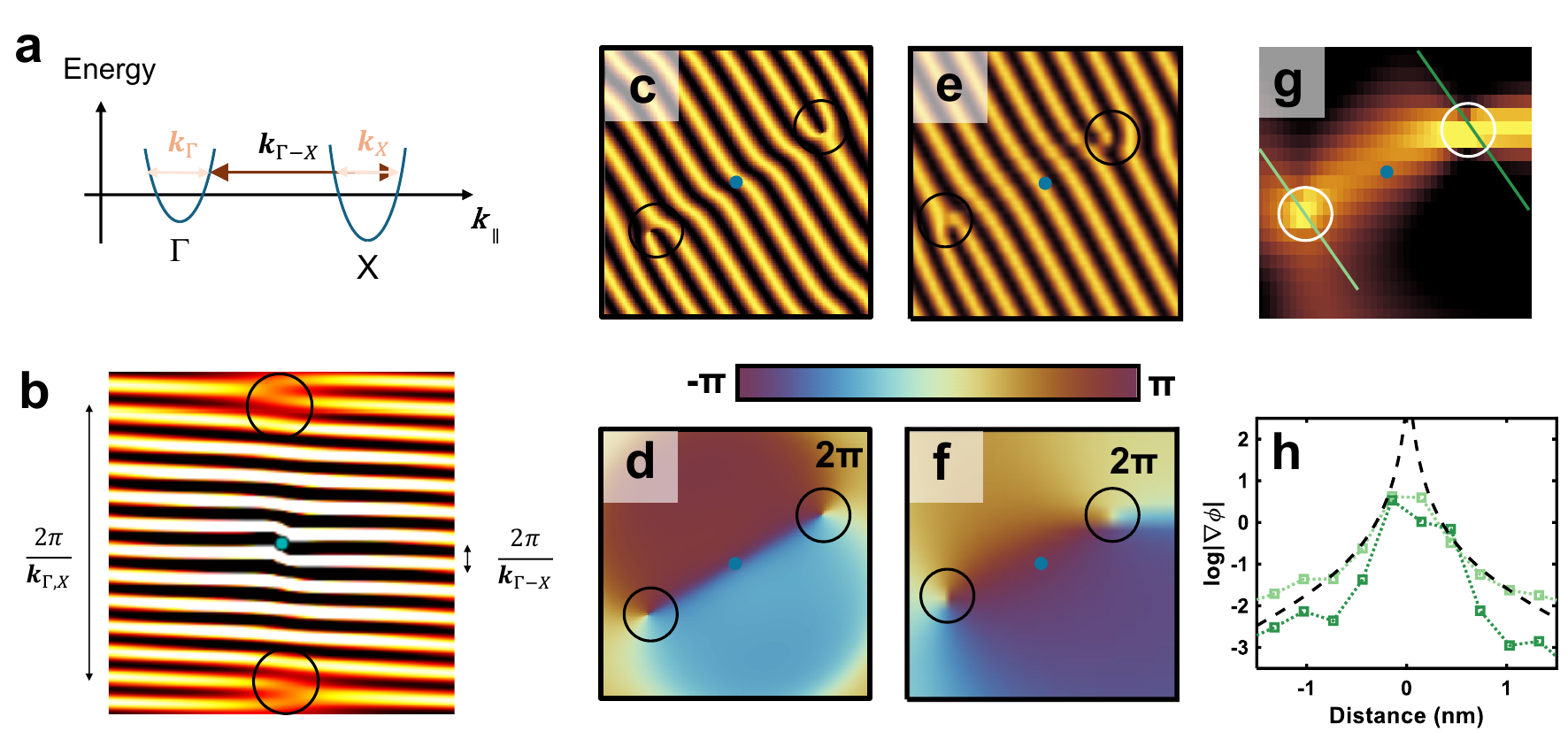}
	\caption{~{\bf{Visualizing scattering textures  induced by SOC in the scattering pattern.}} {\bf a}~Schematic illustration of two small bands located at the center and at an edge of a square Brillouin zone in a two-dimensional electron gas. This leads to two well defined DOS patterns, shown in {\bf b}. One is energy independent and provides small wavelength oscillations in the LDOS at $2\pi/{\bf k}_{\Gamma-X}$. The other one, ${\bf k}_{\Gamma,X}$, leads to large wavelength energy-dependent oscillations. When considering SOC, we find phase slips located on the large wavelength oscillations at $2\pi/{\bf k}_{\Gamma,X}$. We provide the LDOS and phase patterns obtained from calculations with realistic band parameters in {\bf c,d}. In {\bf e,f} we show experimentally observed scattering and phase patterns close to an impurity. In {\bf g} we show a map of the absolute value of the gradient of the phase and in {\bf h} cuts of this quantity (squares) along the light and dark green lines in {\bf g}. The same calculated quantity is shown as a dashed line in  {\bf h}. The phase slips are marked by circles. Impurities are shown as blue disks.}
	\label{FigureScheme}
\end{figure*}

To explore spin-orbit scattering with detail, we consider a model consisting of a single impurity with spin–orbit coupling and two circular bands on one of the sides of a square lattice (Fig.~\ref{FigureScheme}{\bf a}). We include Bragg scattering between bands crossing the Fermi level at a momentum separation matching that of the experimentally observed one-dimensional modulation. The impurity is placed at the center of a nanoscale region and the resulting local density of states is computed (Methods Section F).

In the absence of interband coupling, intraband scattering produces circular interference patterns with wavelength $2\pi/k_{\Gamma,X}$ in each band, while the Bragg component generates a shorter-wavelength modulation set by $2\pi/k_{\Gamma - X}$. When spin–orbit scattering is included, interference between the two bands gives rise to phase dislocations located at a radius $\sim 2\pi/k_{\Gamma,X}$ along the direction perpendicular to the primary modulation direction (Fig.~\ref{FigureScheme}{\bf b}, further details in Extended Data Figure \ref{SuplModel1}). The position of the phase slips evolves with energy, reflecting the dispersion of the underlying bands through the energy dependence of $k_{\Gamma,X}$. As we discuss in Methods Section F, their spatial structure is largely insensitive to the relative position of band extrema with respect to the Fermi level, as well as to the electron or hole character and band curvature. The primary feature is that the phase slips appear always in pairs at the radius given by $\sim 2\pi/k_{\Gamma,X}$.  We provide calculations using parameters described in Methods Section F in Fig.~\ref{FigureScheme}{\bf c}, {\bf d}. An example for the features found around an impurity site is shown in Fig.~\ref{FigureScheme}{\bf e}, {\bf f}.

To quantify the local influence of spin–orbit scattering, we analyze the spatial dependence of the phase gradient, $|\nabla \phi(\mathbf{r})|$ (the spatial map is shown in Fig.~\ref{FigureScheme}{\bf g}, for the same field of view as Fig.~\ref{FigureScheme}{\bf e}). Near the dislocation core, the gradient decays approximately exponentially, $|\nabla \phi(d)| \propto e^{-d/d_0}$, as we show by the dashed black line in in Fig.~\ref{FigureScheme}{\bf h}, where $d$ denotes the distance from the phase slip center. The associated decay length $d_0$ increases monotonically with the relative strength of spin–orbit to scalar scattering, yielding $d_0 \propto t_{\mathrm{SOC}}/t_0$ (with $t_{0}$ the strength of scalar scattering and $t_{\mathrm{SOC}}$ the strength of spin orbit scattering, see Methods Section F and Extended Data Figure \ref{SuplModel3}). Fitting the experimental data (Fig.~\ref{FigureScheme}{\bf h}) to an exponential decay, $|\nabla \phi(d)| \propto e^{-d/d_0}$, yields a characteristic length scale consistent with a ratio of spin–orbit to scalar scattering strengths of $t_{\mathrm{SOC}}/t_0 \approx 0.8$.

\paragraph*{Discussion.}

While our model reproduces the emergence of phase slips and captures key features observed in several regions of the experimental maps, the measurements often reveal considerably more intricate patterns, particularly in areas with closely spaced impurities. In these regions, phase slip positions exhibit a strong dependence on bias voltage, and are distributed along multiple directions rather than being confined perpendicular to the primary one-dimensional modulation, as shown in Fig.~\ref{FigureScheme}. This behavior points to a more complex electronic landscape, in which multiple bands contribute near the Fermi level and the local chemical potential varies significantly across the sample. As a result, the position and structure of phase slip centers are modified by spatial inhomogeneity. As discussed in the Supplementary Information Section II, impurity-induced localized states interact within charge puddles and can be described in terms of potential wells arising from variations in the local chemical potential. The application of a magnetic field further increases the density of states and leads to similar patterns. Nevertheless, the model produces phase slips with strongly varying band structure parameters (Fermi level as well as effective mass) and the internal structure of the scattering potential changes the direction of the position of phase slips (Methods Section F and Extended Data Figures \ref{SuplModel2},\ref{ChargeDensity}). The overall robustness of the observed phase singularities, together with their consistency with the model, indicates that spin–orbit scattering constitutes an important mechanism for the formation of phase slips. In particular, the emergence of well-defined dislocations requires spin–orbit and scalar scattering amplitudes to be comparable in magnitude, $t_{\mathrm{SO}} \sim t_0$.

These results demonstrate that the identification of a dominant wavevector, enabling the selection of a well-defined and localized region in reciprocal space, provides a powerful route to extract phase information from tunneling conductance measurements and the local density of states. In the present case, the $2 \times 1$ surface reconstruction, which gives rise to one-dimensional Eu rows, is essential to isolate the modulation and reveal the associated phase winding around Eu interstitials. More generally, any well-defined wavevector producing an energy-independent spatial modulation of the DOS should allow a similar phase reconstruction. Natural candidates in other quantum materials include surface reconstructions, Bragg wavevectors of the underlying lattice, or collective modulations such as charge density waves. The observation and spatial characterization of phase slips thus provide direct access to the role of spin–orbit interactions at scattering centers, offering a level of insight that is difficult to achieve through conventional approaches.

\paragraph*{Conclusion.}

We have shown that Eu interstitials in EuCd$_2$As$_2$ create a landscape of localized states and phase singularities associated with spin-orbit scattering. Since geometric phase governs key electronic properties—including spin precession, spin relaxation, charge transport and the stability of spin textures in systems with spin–orbit coupling\cite{Manchon2015RashbaReview,RevModPhys.76.323,Sinova2015}—the ability to visualize these defect-induced phase textures provides a new route for understanding and engineering electronic states in magnetic semiconductors and related quantum materials.

\clearpage
\newpage

\clearpage
\newpage


\clearpage
\pagebreak

\setcounter{figure}{0}

\renewcommand{\thefigure}{ED\arabic{figure} }

\section*{M\lowercase{ethods}}

\subsection*{The compound EuCd$_2$As$_2$}

The compound EuCd$_2$As$_2$ is a magnetic semiconductor with a band gap ranging from 0.77 eV, measured from optical measurements\cite{PhysRevLett.131.186704} to about 0.05~eV \cite{shi2023absence} obtained in some samples from resistivity vs temperature measurements. Anomalous Hall conductivity and Nernst effect, commonly associated to SOC\cite{Nagaosa2010,Sinova2015}, were reported\cite{PhysRevLett.126.076602,roychowdhury2023,PhysRevB.94.045112,PhysRevB.97.214422}.  This system was previously proposed to be a magnetic Weyl semimetal presenting a single pair of Weyl points\cite{li2022,soh2019,cao2022,valadkhani2023,wang2019,ma2019,ma2020}. However, subsequent work unveiled that this material is instead a semiconductor with a subtle band structure which favors formation of excitations close to the Fermi level\cite{PhysRevLett.131.186704,Kuthanazhi,PhysRevMaterials.7.034402,shi2023absence,nishihaya2024intrinsic}. These are, a vanishing or small conductance close to the Fermi level, presence of localized states and charge puddles, and LDOS patterns with a pronounced multifractal character. All these results, shown in the Supplementary Information, are comparable to those obtained in other doped semiconductors\cite{MORGENSTERN20121795,PhysRevB.80.161311,doi:10.1126/science.aag1715,10.1063/1.4942517,PhysRevB.92.085140,Morgenstern2012}.

Europium is in a divalent Eu$^{2+}$ (4f$^7$) configuration resulting in no f-states close to the Fermi level. The content of Eu in samples considerably influences its properties\,\cite{PhysRevLett.131.186704,shi2023absence,PhysRevMaterials.7.034402,nahyun2020}, with samples being antiferromagnetic (AFM) with T$_N$ = 9 K or ferromagnetic (FM) with T$_C$ = 26~K\,\cite{nahyun2020,gati2021, yu2023}. Here we measure an AFM sample, as described below.

\subsection*{S\lowercase{urface termination and topographic features in} E\lowercase{u}C\lowercase{d}$_2$A\lowercase{s}$_2$}

The crystal structure of EuCd$_2$As$_2$ belongs to the space group $P\overline{3}m1$ with lattice parameters $a=b$=~4.45~\AA~and $c$=~7.38~\AA\cite{zaac.201100179,zaac.19966220418}. The atomic arrangement consists of layers of Eu atoms separated by As-Cd-Cd-As layers. Bonds between As and Cd are strong, so that the material cleaves leaving exposed Eu or As surfaces, both having a hexagonal shape.

On the surface we most often observe the row-like features discussed in the main text. The atomic Eu-Eu distance within each row is of 4.5$\pm0.1$~\AA, coinciding with the distance between Eu atoms in EuCd$_2$As$_2$. The distance between rows is $d=7.8 \pm0.1$~\AA, which is $d=2a \cos(30^{\circ})$, with $a$=~4.5~\AA. The rows are thus a 2$\times$1 reconstruction of the hexagonal Eu surface lattice.

The one-dimensional row-like atomic arrangements observed in our STM topography images (Fig.\,\ref{FigureTopo}{\bf c}) reminds the 2$\times$1 missing row reconstruction often found in transition metal compounds and in other pnictides\cite{PhysRevLett.59.1833,yin2009,fente2018,Hoffman_2011}. For instance, in 5d metals such as Au, the $2\times1$ reconstruction appears because of the interplay between d-electron hybridization and Smoluchowski smoothing of the s-electron surface charge\cite{Kern1993,MWFinnis1974}. The number of nearest neighbors in the reconstructed surface is the same as in the unreconstructed surface. However, there is more space to spread the charge and reduce the contribution to the kinetic energy of surface electrons\cite{PhysRevLett.59.1833}. Similarly, Eu in the $2\times1$ reconstructed surface of EuCd$_2$As$_2$ has three As nearest neighbors, as in a full hexagonal Eu layer. This allows electrons derived from other orbitals to spread, possibly favoring the downward shift of empty states observed in DFT calculations (from Fig.\,\ref{Figure2}{\bf a} to Fig.\,\ref{Figure2}{\bf c}).

\begin{figure}
	\includegraphics[width=1\columnwidth]{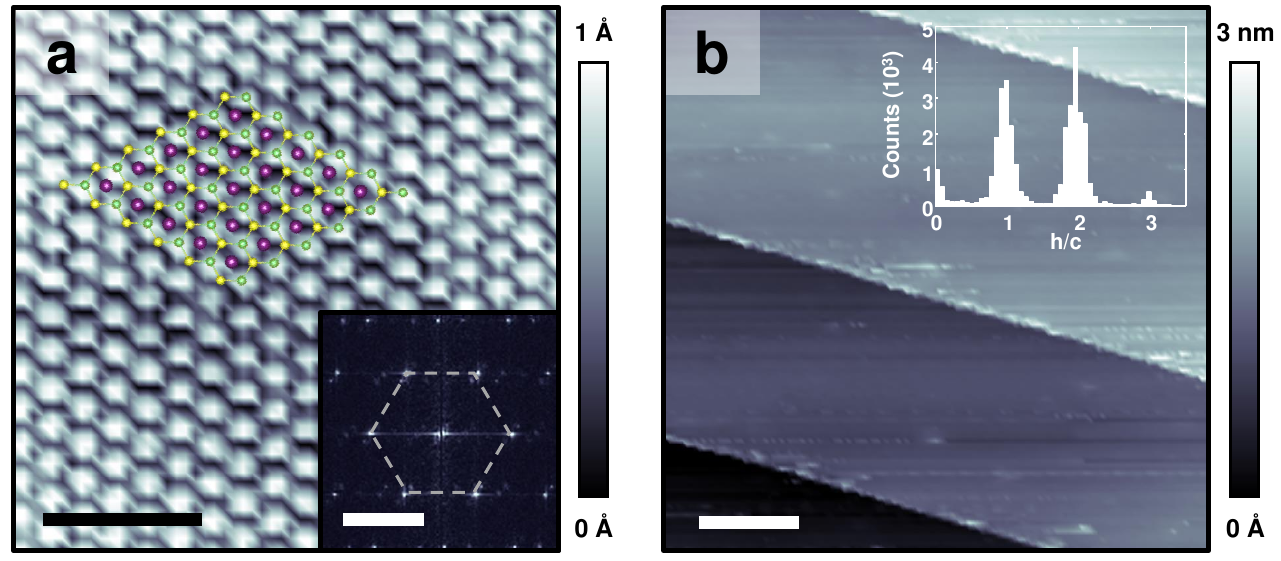}
	\caption{{\bf a} Atomic resolution STM topographic image measured at $+$100~mV and 0.2~nA. These surfaces are much less frequent than the one containing rows discussed in the publication. The hexagonal atomic structure with the atomic lattice is superimposed (colored circles, color code follows the Fig.\,1 of the main text, violet for Eu, green for As and yellow for Cd). Horizontal scale bar is 1 nm long. In the inset we show the Fourier transform of the main panel (horizontal scale bar is 3~nm$^{-1}$ long). The white hexagon shows the Bragg peaks of the hexagonal atomic arrangement seen in the main panel. {\bf b} Large scale topographic STM image measured at 100 mV and 0.4 nA. Horizontal scale line is 40 nm long. In the inset we show the height histogram of the image, with the distance normalized to the $c$-axis lattice constant.}
	\label{SuplLattice}
\end{figure}

The surface discussed in the main text is the one most frequently observed. Nevertheless, we could also find the hexagonal lattice, possibly formed by an As surface, similar to the one studied in Refs.\,\cite{roychowdhury2023,Kong2025}. In Fig.\,\ref{SuplLattice}{\bf a} we provide a topographic image of this surface, with the atomic lattice superposed. The tunneling conductance on this surface is qualitatively similar to the one obtained on the other surfaces, although we do not observe the charge puddles discussed in the main text, suggesting a more metallic-like character on this termination. Generally, atomically flat areas are separated by steps. We show some steps in Fig.\,\ref{SuplLattice}{\bf b}. From the height histograms of such images (inset of Fig.\,\ref{SuplLattice}{\bf b}) we find that steps are almost always separated by the c-axis lattice distance.

\subsection*{STM measurements and sample synthesis}

We performed STM measurements using a home-made STM described in Refs.\,\cite{suderow2011,marta2021}, operating at 4.2~K, and the software described in Ref.\,\cite{fran2021}. The samples were grown by excess of Sn flux as described in Refs.\,\cite{gati2021,PhysRevMaterials.7.034402,shi2023absence,Kuthanazhi}. We use very high purity Eu, as in Refs.\,\cite{gati2021,Kuthanazhi,PhysRevMaterials.7.034402}. Our samples present an A-type AFM order below $T_N$ = 9.5 K \cite{nahyun2020}, as in similar samples\,\cite{PhysRevLett.131.186704,Kuthanazhi,PhysRevMaterials.7.034402,shi2023absence}.

Before measuring, we cleaved the samples \textit{in situ} in the (001) plane at 4~K using the method described in Ref.\,\cite{suderow2011}, obtaining clean and atomically flat surfaces. We cleaved 4 different samples, 15 times in total. To analyze data we used the methods described in Refs.\,\cite{fran2021, horcas2007}. To obtain the multifractal properties of the tunneling conductance maps, discussed in the Supplementary Information, we have followed the method described in Refs.\,\cite{pepe2020, postolova2020,morgenstern2003, terao1996}.

\subsection*{DFT calculations}

Electronic band structure calculations within DFT\,\cite{Hohenberg1964,Kohn1965} were made by using the pseudo-potential augmented plane-wave\,\cite{Bloechl1994,Kresse1999} (PAW) basis set implemented in the Vienna Ab initio Simulation Package (VASP)\,\cite{Kresse1993,Kresse1996,Kresse1996_2}.
All calculations were performed with the Perdew–Burke–Ernzerhof~\cite{Perdew1996} generalized gradient approximation (GGA), and a plane-wave cutoff of 500\,eV was used. 
First the bulk structure was relaxed on a $ 6\times6\times4$ $k$-point mesh using the primitive unit cell of EuCd$_2$As$_2$.
To account for the $2\times1$ row-like Eu structure found at the surface of STM, we have transformed the relaxed unit cell to a monoclinic $\sqrt{3}\times1\times1$ unit cell. We considered a $k$-mesh of $10\times15\times1$ for our slab calculations consisting of one monolayer. This provides us with two inequivalent Eu sites in the $ab$ plane and we could remove every second row of Eu from the surface layer. To obtain the band structure at the surface, we used double Eu terminated slabs with two CdAs layers in between. For the results around Eu interstitials we introduced an interstitial and further expanded the monoclinic unit cell to a $2\times2\times1$ cell also as a monolayer slab. We then relaxed again, using a 15 \AA~void on the surface. We focus on simulations done by calculating the charge density in a very narrow energy range around the Fermi level of $-0.1$ to $+0.1$ eV, to compare as closely as possible with STM experiments.

To understand the role of the 4f Eu states, we first performed bulk GGA+U calculations with U = 5 eV. This places the 4f states deep in the valence band, at energy locations similar to those reported by ARPES measurements\,\cite{ma2019,ma2020}. From DFT calculations, in absence of SOC, the electronic structure of  EuCd$_2$As$_2$ exhibits an energy gap of $E_{\Delta}\approx60$ meV. Upon inclusion of SOC, the system becomes weakly metallic, characterized by a very low density of states at the Fermi level\cite{valadkhani2023}. To reduce the computational cost of the monolayer calculations, the Eu ions were constrained to the $\text{Eu}^{2+}$ valence state by excluding the $4f$ states from the valence manifold. This approximation has been shown to provide a reliable description of the low-energy electronic structure\cite{song2023}. Within this framework, the monolayer exhibits a quantum-confinement-induced gap of $ \approx 222$~meV. This resembles the one found recently in experiments in EuCd$_2$As$_2$\,\cite{PhysRevLett.131.186704,shi2023absence}. The electronic structure near the Fermi level remains largely unchanged relative to the nonrelativistic GGA+$U$ calculations, except for a slightly larger energy gap. The magnetocrystalline anisotropy favors out-of-plane spin alignment, with a magnetocrystalline anisotropy energy  of approximately $50$~$\mu\text{eV}$.

\subsection*{P\lowercase{hase patterns at the} E\lowercase{u rows}}

\begin{figure}
	\includegraphics[width=1\columnwidth]{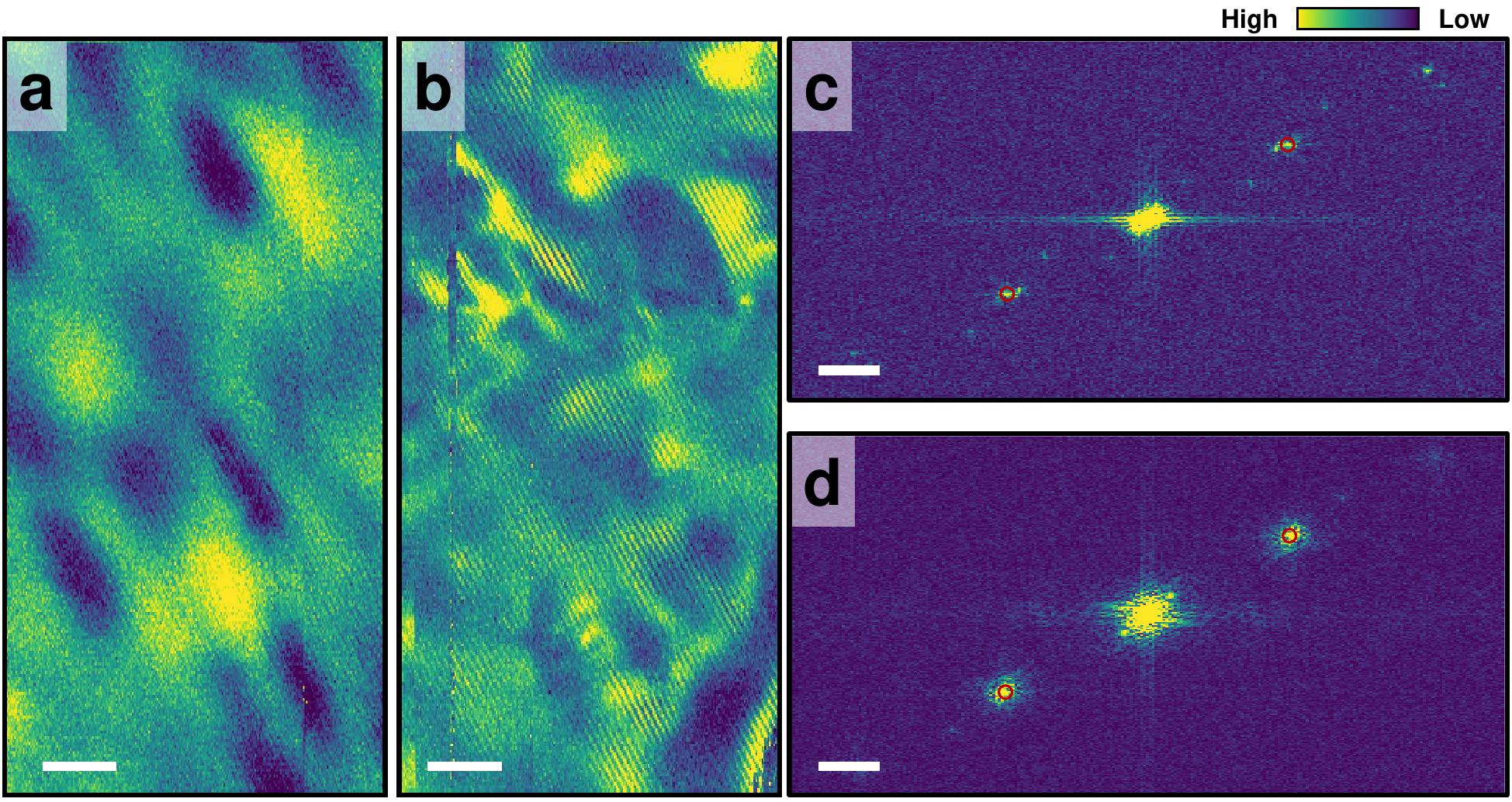}
	\caption{{\bf a} Tunneling conductance map at zero bias and zero magnetic field. {\bf b} Tunneling conductance map at zero bias and 14~T. {\bf c} Fourier transform of {\bf a}. {\bf d} Fourier transform of {\bf b}. We highlight with red circles the one-dimensional modulation we have used to calculate the phase maps shown in Figs.\,\ref{SuplPhaseFullMaps0T},\ref{SuplPhaseFullMaps14T}. White scale bars are 10~nm long in {\bf a}, {\bf b} and  1~nm$^{-1}$ long in {\bf c}, {\bf d}.}
	\label{SuplExampleFilter}
\end{figure}

To obtain the phase pattern, we start by filtering out the one-dimensional features due to the Eu rows. We show in Fig.\,\ref{SuplExampleFilter} two maps of the tunneling conductance presenting a one-dimensional feature at a wavevector $\textbf{Q}_{1D}$. We can write quite generally for such maps:

\begin{equation}
T(\textbf{r})= A_{1D}(\textbf{r}) \, e^{i \, \textbf{Q}_{1D} \cdot \textbf{r}} +...,
\label{topoec}
\end{equation}
where $\textbf{r}=(x,y)$ is a real space vector, $A_{1D}$ the amplitude of the modulation and ... represents all other contributions to the map\,\cite{Lawler2010,Herrera2021}. We multiply by $e^{i \, \textbf{Q}_{1D}}$ and shift in reciprocal space the origin to the first peak of the one dimensional modulation. We then make a Gaussian filter and obtain
\begin{equation}
A_{1D}(\textbf{r})= \frac{1}{\sqrt{2 \pi} \sigma} \, \int d\textbf{R} \, T(\textbf{R}) \, e^{i \, \textbf{Q}_{1D} \cdot \textbf{R}} \, e^{-\frac{{\left( \textbf{r} - \textbf{R} \right)}^2}{2 \sigma_r^2}},
\label{lockinreal}
\end{equation}
where $\sigma_r$ is the cutoff length of the Gaussian filter in real space. $\sigma_r$ should be such that the one-dimensional row is included, but that other Bragg peaks are excluded from the analysis. We can compute $A_{1D}$ in reciprocal space and then Fourier transform back to real space, 
\begin{equation}
\begin{split}
A_{1D}(\textbf{r})= & F^{-1}[A_{1D}(\textbf{q})]= \\ = &  F^{-1}\left[ F \left(  T(\textbf{r}) \,  e^{i \, \textbf{Q}_{1D} \cdot \textbf{r}}    \right) \cdot  \frac{1}{\sqrt{2 \pi} \sigma}  e^{-\frac{\textbf{q}^2}{2 \sigma_q^2}}  \right],
\label{lockinfft}
\end{split}
\end{equation}
with $\sigma_q=1/\sigma_r$ and  $F[ \, ]$ representing the Fourier transform.

We can now make maps of the phase of the one-dimensional modulation, by plotting $\phi_{1D}(\textbf{r})=arctan\left(  \frac{Im(A_{1D}(\textbf{r}))}{Re(A_{1D}(\textbf{r}))}     \right)$ as a function of the position $\textbf{r}=(x,y)$.

\begin{figure}
	\includegraphics[width=1\columnwidth]{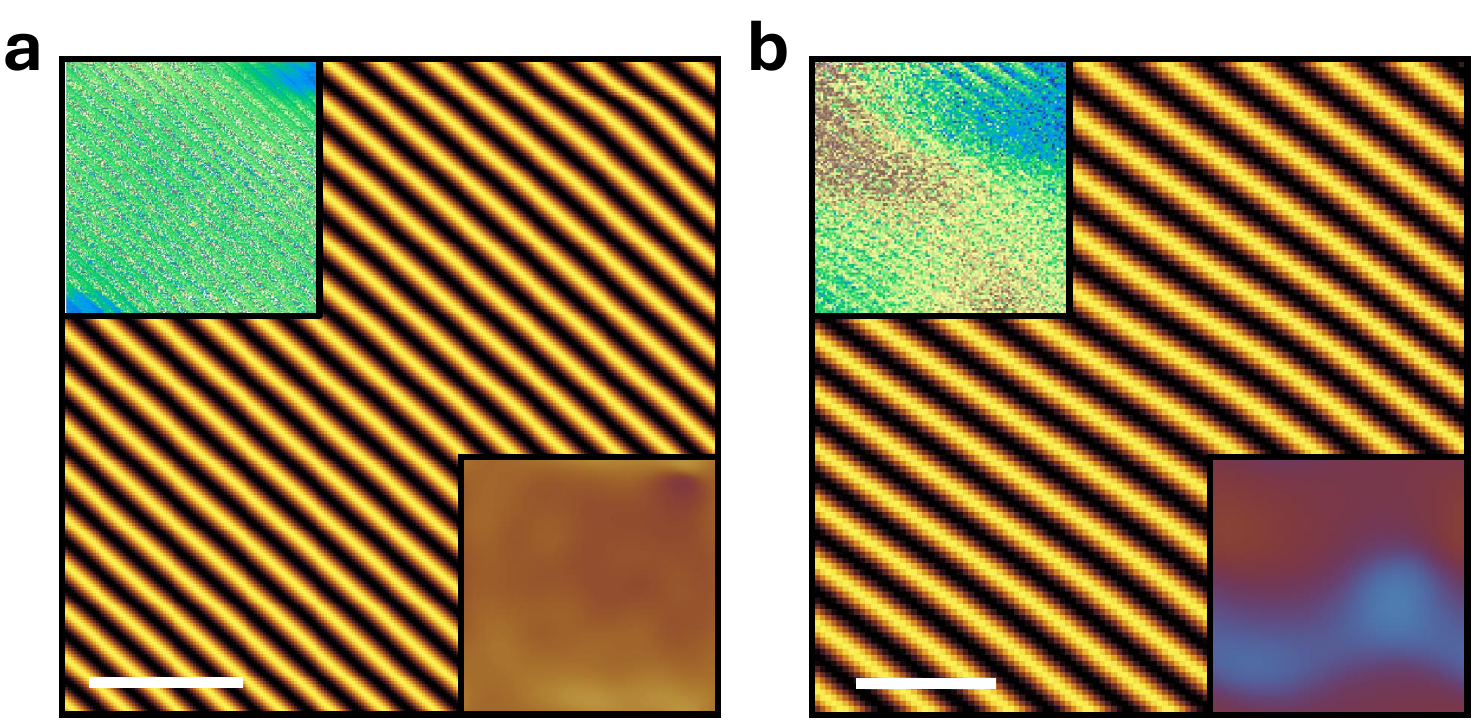}
	\caption{Real space image obtained from the phase map (lower inset) at areas without impurities at zero field ({\bf a}) and 14 T ({\bf b}). Horizontal scale line is 5 nm long. In the upper inset, conductance map at 0 mV in each field of view.}
	\label{MapsNoImp}
\end{figure}

The result for areas with impurities is discussed in the main text. For areas without impurities, we show the result in Fig.\,\ref{MapsNoImp}. We do not observe phase slips in these areas. We show in Fig.\,\ref{GaussianComp} the result of modifying the cutoff of the Gaussian filter. When the Gaussian filter is too large in reciprocal space, see Fig.\,\ref{GaussianComp}{\bf a}, we find contributions from other wavevectors, and when it is too small in reciprocal space, Fig.\,\ref{GaussianComp}{\bf e}, we find results only from part of the impurities. For all other figures we take a radius of 0.33 nm$^{-1}$.

\begin{figure*}
\begin{center}
	\includegraphics[width=0.85\textwidth]{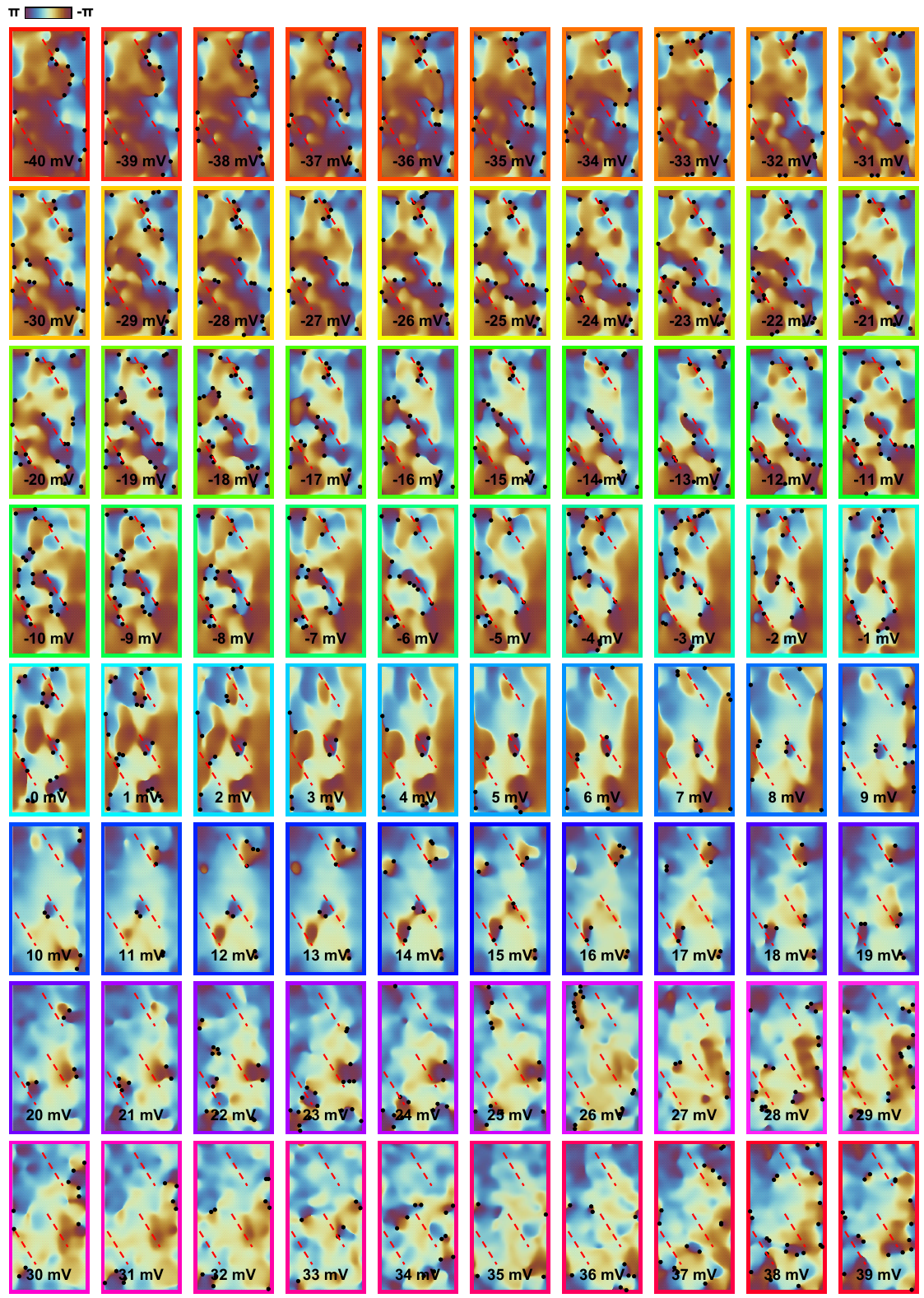}
\end{center}
	\caption{Phase maps at zero magnetic field in the same field of view as Fig.\ref{SuplExampleFilter}({\bf a}). Dashed red lines provide positions of a few impurities, and are discussed in the Supplementary Information Section~II.}
	\label{SuplPhaseFullMaps0T}
\end{figure*}

\begin{figure*}
\begin{center}
	\includegraphics[width=0.85\textwidth]{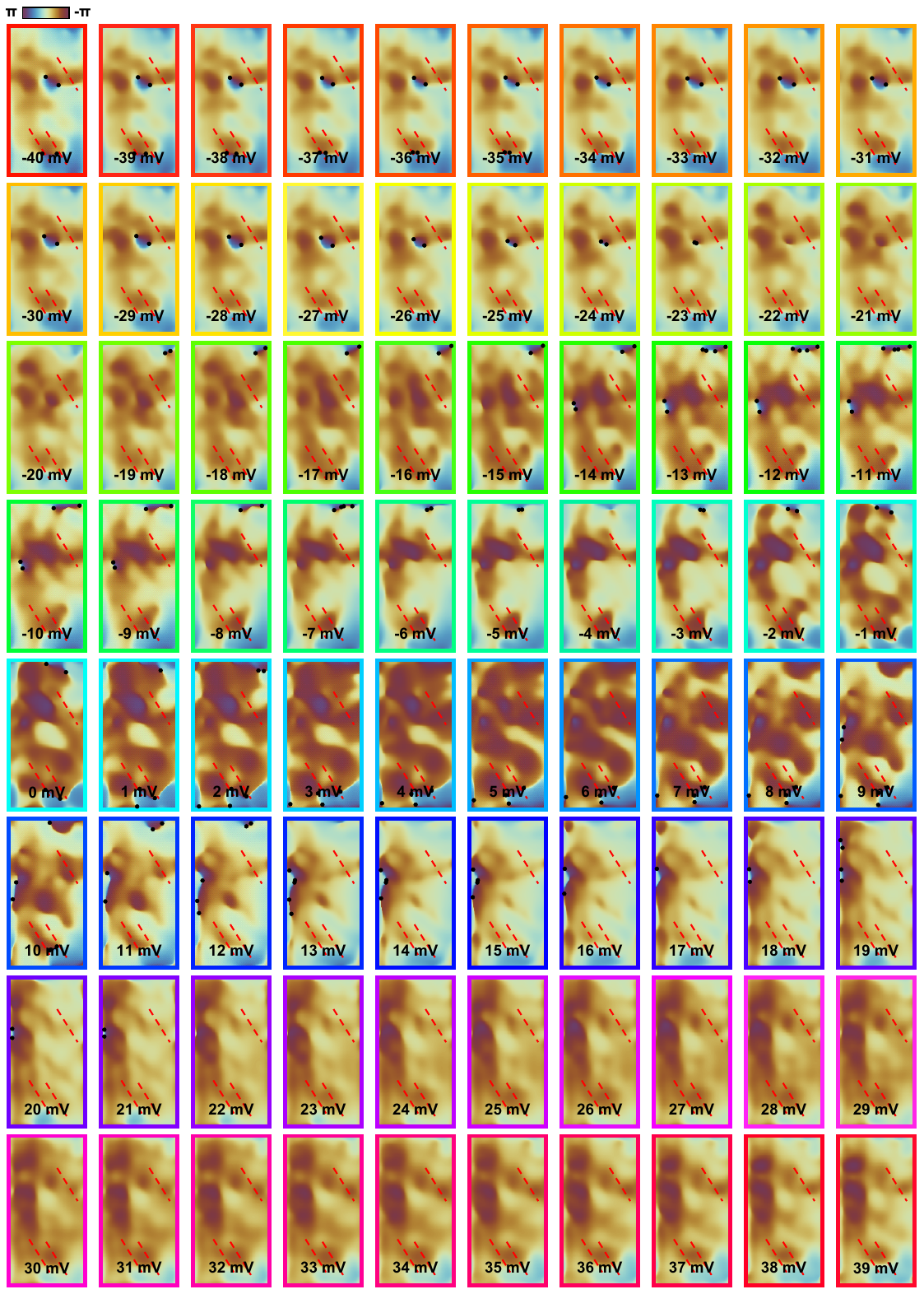}
	\end{center}
	\caption{Phase maps at 14~T in the same field of view as Fig.\ref{SuplExampleFilter}({\bf b}). Dashed red lines provide positions of a few impurities, and are discussed in the Supplementary Information Section~II.}
	\label{SuplPhaseFullMaps14T}
\end{figure*}

For completeness, we provide the phase over two different fields of view in Fig.\,\ref{SuplPhaseFullMaps0T} at zero field and in Fig.\,\ref{SuplPhaseFullMaps14T} for a field of 14~T. The phase slip positions strongly depend on the bias voltage. In Fig.\,\ref{SuplPhaseFullMaps0T} we can identify a significant number of phase slips, with $2\pi$ phase winding at some locations. The number of $2\pi$ phase windings tends to reduce at positive bias voltages when there are less localized states, suggesting a smaller two-dimensional electronic density. The result under magnetic fields suggests that the multiplicity of states increase the overlap between electronic states on different locations, providing only smooth changes in the one-dimensional charge modulation, with a smaller number of phase slips.

\subsection*{C\lowercase{alculations of} SOC \lowercase{scattering}}

\subsection*{Scattering model}

We consider interband scattering mediated by the discrete impurity levels \cite{Dutreix2019, Bacsi2010, Zhang2021}, which is described by the coupling Hamiltonian between the two systems, given by

\begin{equation}
H_I = \sum_{\mathbf{k}\sigma m s}
\left(
t_{\mathbf{k}\sigma;ms}\, c^\dagger_{\mathbf{k}\sigma} c_{m s}
+
t_{\mathbf{k}\sigma;ms}^\ast\, c^\dagger_{m s} c_{\mathbf{k}\sigma}
\right),
\end{equation}
where $c_{m s}$ annihilates an electron in the impurity state $m$ with spin $s$, and
$c_{\mathbf{k}\sigma}$ annihilates an electron with wavevector $\mathbf{k}$ and spin $\sigma$.
The quantity $t_{\mathbf{k}\sigma;ms}$ is the hopping amplitude between the propagating and impurity states.

The variation of the Green’s function describing scattering from a state $(J,\mathbf{k},\sigma)$
to a state $(J',\mathbf{k}',\sigma')$ is
\begin{equation}
\delta \hat{G}_{J\mathbf{k};J'\mathbf{k}'} =
\hat{g}_{J\mathbf{k}}\,
\hat{V}_{J\mathbf{k};J'\mathbf{k}'}\,
\hat{g}_{J'\mathbf{k}'} .
\end{equation}

A hat denotes a matrix in the spin basis. The Green’s function $\hat{g}_{J\mathbf{k}}$ is expanded around the high-symmetry point $J$ using the $\mathbf{k}\!\cdot\!\mathbf{p}$ approximation, $\mathbf{k}\to \mathbf{J}+\mathbf{k}$.
The self-energy is
\begin{equation}
\hat{V}_{J\mathbf{k};J'\mathbf{k}'} = t_0^2\hat{S}^{0}_{JJ'} + t_{SO}^2\, (\mathbf{k}\times\mathbf{k}')_{z}\hat{S}^{1}_{JJ'},
\end{equation}
where $S^{0,1}_{JJ'}=p^{0,1}_{J}G_{imp}p^{0,1}_{J'}$ are matrices encoding the scattering between the states and the impurity orbitals, with $\hat{G}_{imp}$ denoting the impurity Green’s function and $p^{0,1}_{J}$ the coupling matrix between the itinerant band J and the impurity levels. The resulting Green’s-function take the form
\begin{equation}
\begin{aligned}
\delta\hat{G}_{J\mathbf{k};J' \mathbf{k}'}
=&\;\hat{g}_{J+\mathbf{k}}\, t_0^2\, \hat{S}^{0}_{JJ'}\, \hat{g}_{J'+\mathbf{k}'} \\
&+\hat{g}_{J+\mathbf{k}}\, t_{SO}^2\,
(k_x k_y' - k_y k_x')\,
\hat{S}^{1}_{JJ'}\,
\hat{g}_{J'+\mathbf{k}'} .
\end{aligned}
\end{equation}

Here, $t_{0}$ denotes the scalar hopping amplitude, while $t_{SO}$ characterizes the spin-orbit scattering \cite{PhysRevB.95.115307,Nagaosa2010,Sinova2015}. We next consider intervalley processes between the $\Gamma$ and $X$ points.
The real-space variation of the Green’s function is
\begin{equation}
\begin{aligned}
\delta \hat{G}(\mathbf{r},\mathbf{r}')
=
e^{i\mathbf{X}\cdot\mathbf{r}}
\!\!\int\! dk\, dk'\,
e^{i( \mathbf k \cdot \mathbf r - \mathbf k'\cdot \mathbf r')}
\hat{G}_{X\mathbf{k}\Gamma \mathbf{k}'}
+\\
e^{i\mathbf{X}\cdot\mathbf{r}}
\!\!\int\! dk\, dk'\,
e^{i( \mathbf k \cdot \mathbf r - \mathbf k' \cdot\mathbf r')}
\hat{G}_{\Gamma \mathbf{k};X\mathbf{k}'} .
\end{aligned}
\end{equation}

The first terms can be written as
\begin{equation}
\hat{g}_X(\mathbf{r})\, \hat{S}^{0}_{X\Gamma}\, \hat{g}_{\Gamma}(\mathbf{r}')
+
\hat{g}_{\Gamma}(\mathbf{r})\, \hat{S}^{0}_{\Gamma X}\, \hat{g}_X(\mathbf{r}'),
\end{equation}
where the real-space Green’s functions are defined by

\begin{equation}
\hat{g}_J(\mathbf{r}) = \int dk\, e^{i(J+\mathbf k)\cdot \mathbf r}\, \hat{g}_{J+\mathbf k},    
\end{equation}

Using a parabolic band approximation with effective mass $m_J$
(related to the band curvature $c_J$),
\begin{equation}
\hat{g}_J(r;\omega)\simeq
\frac{2m_{J}}{\hbar}\,
H_0(\kappa_{J}r)\, \hat{\sigma}_0 .
\end{equation}
with
\begin{equation}
\kappa_J = \sqrt{\frac{1}{c_J}(E - E_J)},
\end{equation}
the radial oscillation wavevector associated with band $J$ where $E_{J}$ is the band minimum and $c_{J}$ 
 is the band curvature. The Fourier transform of the momentum-dependent terms yields
\begin{equation}
\begin{aligned}
\left[
\partial_x \hat{g}_{0X}(r)\, \hat{S}^{1}_{X\Gamma}\, \partial_{y'} \hat{g}_{0\Gamma}(r')
-
\partial_y \hat{g}_{0X}(r)\, \hat{S}^{1}_{X\Gamma}\, \partial_{x'} \hat{g}_{0\Gamma}(r')
\right]
=\\
i\pi A_{\Gamma} A_X e^{i\mathbf{X}\cdot r}
H_1(\kappa_{\Gamma} r) H_0(\kappa_X r)
\sin\theta\,
\hat{S}_X .
\end{aligned}
\end{equation}

Considering an impurity with two spin states, the total correction to the Green’s function becomes
\begin{equation}
\begin{aligned}
\operatorname{Tr}(\delta G(r))
=&\; A_1 H_0(\kappa_{\Gamma} r) H_0(\kappa_X r)\cos(\pi x) \\
&+ A_2 H_1(\kappa_{\Gamma} r) H_0(\kappa_X r)\sin(\pi x)\sin\theta .
\end{aligned}
\end{equation}
where
\begin{align}
A_1 &=
\alpha t^{2}_{0}(p_{\Gamma\uparrow} g_{\uparrow\uparrow} p_{X\uparrow}
+ p_{\Gamma\downarrow} g_{\downarrow\downarrow} p_{X\downarrow}), \\
A_2 &=
\alpha \pi t^{2}_{SO}\,
\big(
p_{\Gamma \uparrow} g_{\uparrow\uparrow} p_{X\uparrow}
-
p_{\Gamma\downarrow} g_{\downarrow\downarrow} p_{X\downarrow}
\big),
\end{align}

and $p_{J,\sigma}$ are the coupling components that encodes the coupling between the bands and the impurity levels with $\alpha = \frac{4 m_X m_{\Gamma}}{\hbar^2}$. Finally, the variation of the electronic density is

\begin{equation}
\delta\rho(r,\theta)
=
A(r)\cos(\pi x)+B(r)\sin(\pi x)\sin\theta,
\end{equation}
with
\begin{align}
A(r) &= \Im\!\left[A_1 H_0(\kappa_{\Gamma} r)H_0(\kappa_X r)\right], \\
B(r) &= \Im\!\left[A_2 H_1(\kappa_{\Gamma} r)H_0(\kappa_X r)\right].
\end{align}

The wavefront dislocations are located at the points where the density modulation vanishes $\delta\rho(r,\theta)=0$. To identify these points, we rewrite the density variation as a complex field in Nye--Berry form

\begin{equation}
\delta\rho(r,\theta) = \Re\!\left[\Phi(r,\theta)e^{i\pi x}\right]
\end{equation}
with

\begin{equation}
\Phi(r,\theta)=A(r)-iB(r)\sin\theta,
\end{equation}
Expressed in polar form with amplitude $\Phi(r,\theta)=|\Phi(r,\theta)|e^{i\phi(\theta,r)}$, the phase of this function is given by
\begin{equation}
\phi(r,\theta) = \pi x + \arg\!\big(\Phi(r,\theta)\big).
\end{equation}

Wavefront dislocations occur where
\begin{equation}
|\Phi(r,\theta)|^2 = A(r)^2 + B(r)^2 \sin^2\theta = 0.
\end{equation}

The condition $A(r)=0$ defines the radius $r_{0}$ where the wavefront dislocations occur
\begin{equation}
A(r_0)=
\Im\!\left[A_1 H_0(\kappa_{\Gamma} r_0)H_0(\kappa_X r_0)\right]=0,
\end{equation}

The vanishing of the second term $B(r)=0$ determines the angular positions of the dislocations for $\sin(\theta)=0$.\\

\begin{figure*}
	\includegraphics[width=1\textwidth]{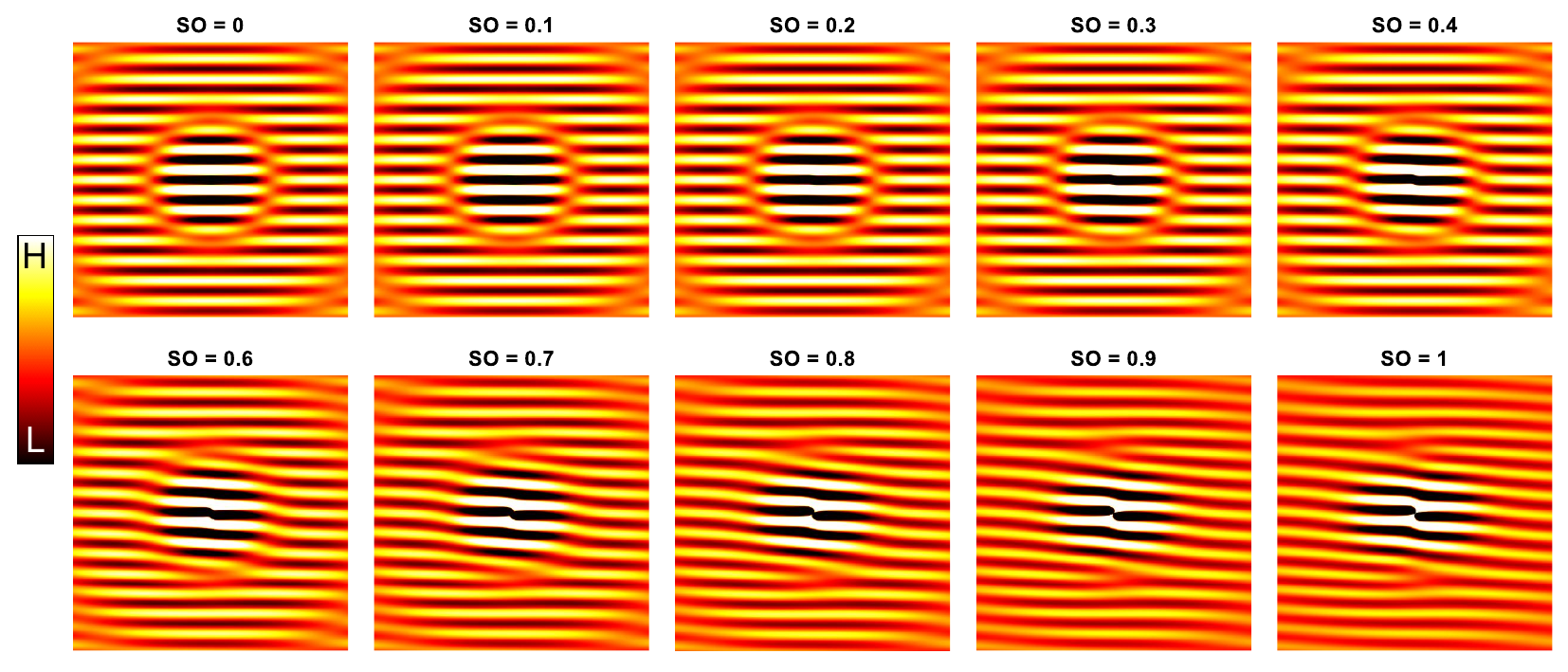}
	\caption{{\bf LDOS maps obtained from two-band scattering as a function of the SOC strength.} We show the LDOS pattern obtained with zero SOC (upper left panel) up to $\frac{t_{SO}}{t_{0}}=1$ (lower right panel). We mark $\frac{t_{SO}}{t_{0}}$ in each panel. The color map follows the bar on the left.}
	\label{SuplModel1}
\end{figure*}

We now analyze LDOS maps for different values of the spin–orbit interaction strength $t_{\mathrm{SO}}$, Fig.\,\ref{SuplModel1}.  For a weak spin–orbit contribution, the isotropic term dominates the density modulation 
\begin{equation}
A(r)=\Im\!\left[t_0^2 A_1 H_0(\kappa_{\Gamma} r)H_0(\kappa_X r)\right]=0,
\end{equation}

The zeros of its amplitude form circular nodal lines that determine the radius $r_{0}$ where the wavefront dislocations emerge (Fig.\,\ref{SuplModel1} upper left panels). As the spin--orbit coupling increases, the vanishing regions of the complex field become localized at $\theta=0$ and $\theta=\pi$ (Fig.\,\ref{SuplModel1} lower right panels). When both contributions are similar, a well-defined dislocation emerges.

\begin{figure*}
	\includegraphics[width=0.9\textwidth]{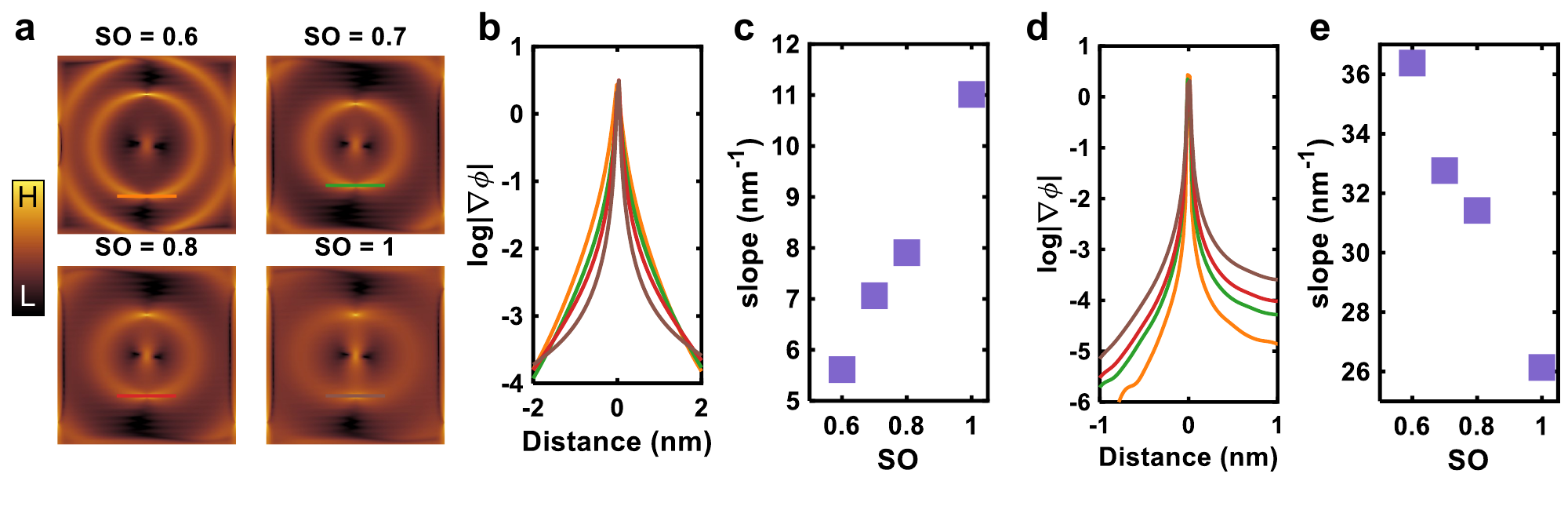}
	\caption{{\bf Dislocation strength $\vert\nabla\phi(\mathbf{r})\vert$ as a function of the SOC scattering strength.} {\bf a} LDOS maps for $t_{SO}/t_0$ between 0.6 and 1. {\bf b} Logarithm of the variation of the phase gradient following the colored lines in {\bf a}. {\bf c} Slope close to the center for the curves in {\bf b}. {\bf d} Logarithm of the variation of the phase gradient perpendicular to the color lines in {\bf a}. {\bf e} Slope close to the center for the curves in  {\bf d}.}
	\label{SuplModel3}
\end{figure*}

We furthermore show $\vert\nabla\phi(\mathbf{r})\vert$ for different SOC scattering strengths in Fig.\,\ref{SuplModel3}. We see in Fig.\,\ref{SuplModel3}{\bf b,d} that $\vert\nabla\phi(\mathbf{r})\vert$ decays very strongly from the center of the phase slip. Close to the center, the decay is roughly exponential, $\vert\nabla \phi (d)\vert\propto e^{-d/d_{0}}$. As we show in Fig.\,\ref{SuplModel3}{\bf c,e}, $d_0$ scales linearly with $t_{SO}/t_0$.

\subsection*{Parameters used to obtain the maps shown in the main text}

We now describe the parameters to obtain the pattern shown in Fig.\,\ref{FigureScheme}{\bf c,d}. The radial oscillations determine the positions of the wavefront dislocations through the wavevector $\kappa_{J}$, which depends on energies measured relative to the Fermi energy $E_{F}$ probed by the STM. To estimate $E_{F}$, we consider the case in which the valence-band edge coincides with $E_{F}$. In this case, the energy spectrum is given by
\begin{equation}
E(k) = \frac{\hbar^2 k^2}{2m_J} - E_F,
\end{equation}
or equivalently
\begin{equation}
E(k) = ck^2 - E_F .
\end{equation}

In this way, a Fermi wavevector can be defined as
\begin{equation}
k_F = \sqrt{\frac{E_F}{c}} .
\end{equation}

We choose a small $k_F$ to ensure the validity of the parabolic band approximation near the Brillouin zone center,
\begin{equation}
k_F \ll \frac{\pi}{a},
\end{equation}
where $a$ is the lattice parameter, with $a = 0.44~\mathrm{nm}$. Thus,
\begin{equation}
\frac{\pi}{a} = \frac{3.141592}{0.44~\mathrm{nm}} = 7.07~\mathrm{nm^{-1}} .
\end{equation}

Defining
\begin{equation}
k_F = \alpha\frac{\pi}{a},
\end{equation}
and taking $\alpha = 0.1$ gives
\begin{equation}
k_F = 0.707~\mathrm{nm^{-1}},
\end{equation}

The wavevector depends on the energy measured relative to the Fermi energy, and the band curvature is
\[
c = 127~\mathrm{meV\,(nm)^2}.
\]
such that a suitable Fermi energy is
\begin{equation}
\begin{aligned}
E_F = c k_F^2 = 127~\mathrm{meV\,(nm)^2}\times(0.707~\mathrm{nm^{-1}})^2 \\
 = 63.64~\mathrm{meV}.
\end{aligned}
\end{equation}

In Fig.\,\ref{FigureScheme}{\bf c,d} we use this $E_F$ and provide the result obtained at $E - E_F = 14~\mathrm{meV}$. With increasing energy, the wavevector increases, leading to a reduction in the radial position of dislocations.

\subsection*{Fermi energies and band curvature}

We now consider a conduction band and a valence band whose
minimum and maximum occur at different energies. The minimum of the conduction band and the maximum of the valence band satisfy the following condition: $E_{\Gamma}<E_{F}<E<E_{X}$. We can write for the radial wavevector

\begin{eqnarray}
\kappa _{\Gamma }=\sqrt{\frac{1}{c_{\Gamma }}(\varepsilon -\Delta _{\Gamma }),%
} \\
\kappa _{X}=\sqrt{\frac{1}{c_{X}}(\Delta _{X}-\varepsilon )},
\end{eqnarray}
with $\varepsilon=E-E_{F}$. Propagating oscillatory solutions are found with real wavevectors. The energy range under consideration lies within $E=[E_{\Gamma },E_{X}]$. Outside this interval, the wavevector becomes imaginary, leading to evanescent states. As a consequence, the interference between the waves responsible for the observed patterns is suppressed, thereby affecting the formation of the oscillatory features.

We define band extrema relative to the Fermi energy because STM measures with respect to zero bias or the Fermi energy. We can write 

\begin{eqnarray*}
\Delta _{\Gamma } &=&E_{\Gamma }-E_{F} \\
\Delta _{X} &=&E_{X}-E_{F},
\end{eqnarray*}
where $\Delta _{\Gamma }$ denotes the position of the conduction-band
minimum relative to the Fermi energy, and $\Delta _{X}$ denotes the position
of the valence-band maximum relative to the Fermi energy.

The radial oscillations are modulated by the Hankel functions associated with each band with wave vector $\kappa_{J}$. The positions of the wavefront dislocations are governed by the band parameters involved in the scattering process, such as the band minima and curvatures. Using
\begin{equation}
H_0(\kappa_{J} r)=J_0(\kappa_{J} r)+iY_0(\kappa_{J} r),
\end{equation}
with
\begin{align}
J_0(\kappa_J r) &\simeq \sqrt{\frac{2}{\pi\kappa_J r}}
\cos\!\left(\kappa_{J} r-\frac{\pi}{4}\right),\\
Y_0(\kappa_J r) &\simeq \sqrt{\frac{2}{\pi\kappa_J r}}
\sin\!\left(\kappa_{J} r-\frac{\pi}{4}\right),
\end{align}
the product becomes
\begin{equation}
\Im\!\left[H_0(\kappa_{\Gamma} r)H_0(\kappa_X r)\right]
=\frac{1}{\pi r\sqrt{\kappa_{\Gamma}\kappa_X}}
\sin\!\left((\kappa_{\Gamma}+\kappa_X)r-\frac{\pi}{2}\right).
\end{equation}

This leads to dislocation positioned at
\begin{equation}
r_0 = \frac{(2n+1)\pi}{2(\kappa_X+\kappa_{\Gamma})} .
\end{equation}

We show in Fig.\,\ref{SuplModel2} results obtained when changing the band edge energies (Fig.\,\ref{SuplModel2}{\bf b}) and effective masses (Fig.\,\ref{SuplModel2}{\bf c}). By changing band curvature and position with respect to the Fermi level, we modify the position of the radial wavefront of the long wavelength modulation, which corresponds to the intraband scattering. This also leads to changes in the position where we observe the dislocations.

\subsection*{Rotated Dislocation Patterns}

The observation of dislocations forming at angles other than perpendicular to the wavefronts suggests that the minimal model does not capture all properties of the observed electronic scattering. We have until now considered the influence of different types and shapes of bandstructures. However, this requires isotropic scattering by point-like impurities, which is rarely obtained in the experiment. We show in Fig.\,\ref{ChargeDensity} DFT calculations of the charge density around an Eu interstitial integrated from $-0.1$ to $+0.1$ eV around the Fermi level. The charge density around the impurity is anisotropic. The anisotropy possibly varies with the chemical potential and with the exact position of the Eu impurity, which might vary from site to site. 

The finite spatial extent of the impurity can be modeled by two closely spaced scattering centres separated by a vector $\mathbf d$. As shown below, this separation gives rise to dislocations rotated with respect to the conventional single scattering center isotropic case. The resulting scattering potential is given by the superposition of the individual center potentials

\begin{equation}
V(\mathbf r)=v(\mathbf r)+v(\mathbf r-\mathbf d).
\end{equation}

For small separations, the scattering potential can be expanded in $\mathbf d$, yielding

\begin{equation}
V(\mathbf r)\approx 2v(\mathbf r)-(\mathbf d \cdot \nabla)v(\mathbf r).
\end{equation}

In reciprocal space, the first-order expansion in the impurity displacement becomes

\begin{equation}
    (\mathbf d \cdot \nabla)v(\mathbf r)\rightarrow i\mathbf d\cdot(\mathbf k-\mathbf k')v(\mathbf k, \mathbf k')
\end{equation}

Neglecting higher-order momentum contributions, the scattering potential is then written as

\begin{equation}
    V_{\sigma,\sigma'}(\mathbf k,\mathbf k')\approx t_{0}(1+i\mathbf d\cdot (\mathbf k-\mathbf k'))S_{J\sigma,J'\sigma'}+t_{SO}(\mathbf k \times \mathbf k')S'_{J\sigma,J'\sigma'}
    \end{equation}

The anisotropic contribution can then be incorporated into the real-space hopping amplitudes, leading to

\begin{equation}
t_{0}^{2}\rightarrow t_{s}^{2}+it_{s}t_{p}\cos (\theta ),    
\end{equation}
where $t_s$ denotes the conventional isotropic scattering amplitude. The angular dependence induced by the finite spatial extent of the impurity is incorporated through an effective parametrization of the hopping amplitude $t_p(\mathbf d)$. Thus, the density variation can be written as

\begin{equation}
\delta\rho(r,\theta)
\approx
\operatorname{Re}
\!\left[
\left(
t_s^2 A'(r)
+
i R(r)\sin(\theta+\delta)
\right)
e^{i\pi x}
\right],
\end{equation}
with

\begin{equation}
R(r)
=
\sqrt{
t_s^2 t_p^2 A'^2(r)
+
t_{\rm SO}^4 B'^2(r)
},
\end{equation}
thereby showing that the anisotropy of the impurity potential induces a
rotation of the phase dislocations by an angle

\begin{equation}
\delta
=
\tan^{-1}
\!\left(
\frac{A'(r_0)t_s t_p}
     {t_{\rm SO}^2 B'(r_0)}
\right).
\end{equation}

\begin{figure}[t!]
	\includegraphics[width=1\columnwidth]{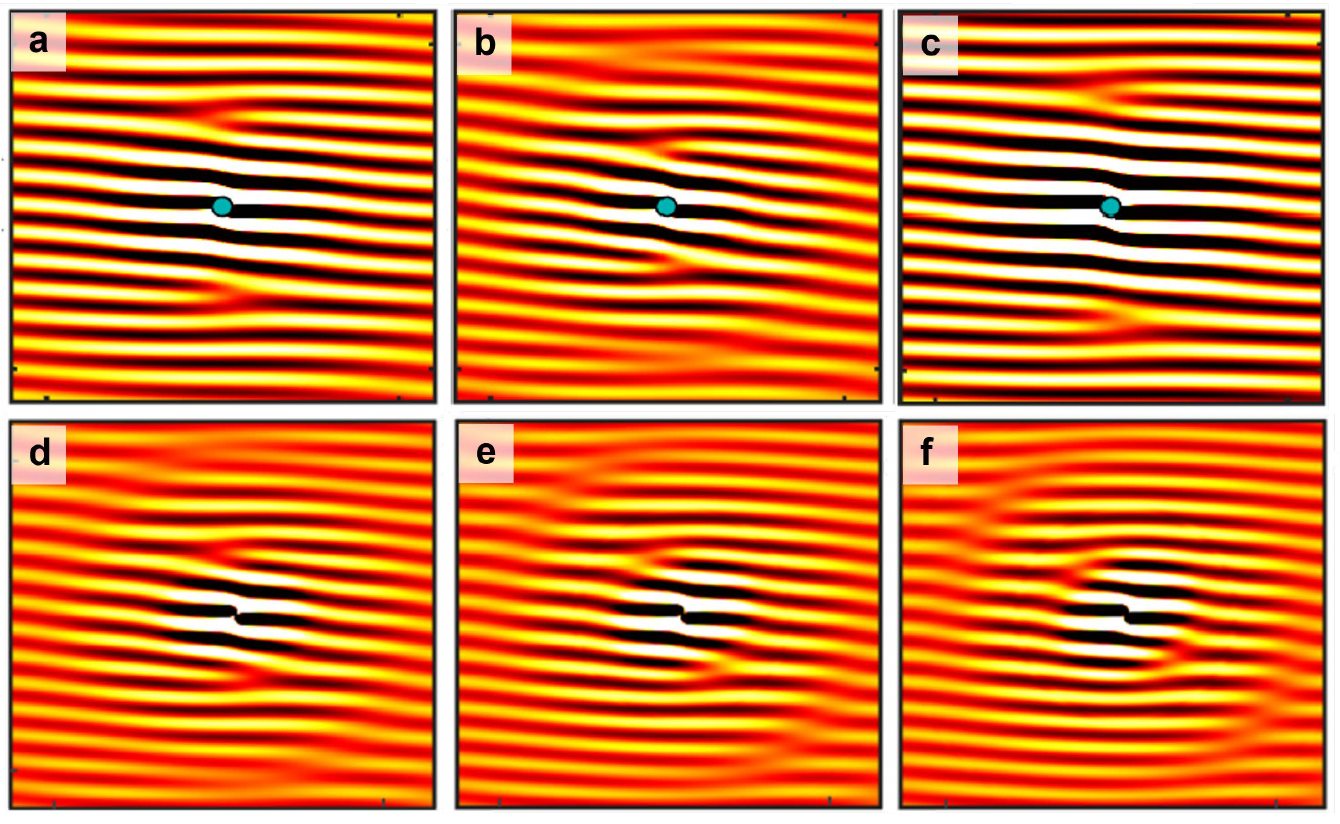}
	\caption{{\bf Varying Band-Edge energies, effective mass and anisotropic hopping amplitude.} In {\bf a} we show the DOS pattern obtained as described in the text. {\bf b} LDOS pattern obtained using $E_X=10$ and $E_{\Gamma}=30$. {\bf c} LDOS pattern obtained using two distinct band curvatures $c_X=3c_{\Gamma}$. In {\bf d} we show the LDOS pattern obtained as described in the text with $t_{p}=0$. LDOS pattern obtained using $t_{p}=0.5$ ({\bf e}) and $t_{p}=0.9$ ({\bf f}). The lateral size of the images is 12 nm.}
	\label{SuplModel2}
\end{figure}

\begin{figure}[t!]
\begin{center}
	\includegraphics[width=0.5\columnwidth]{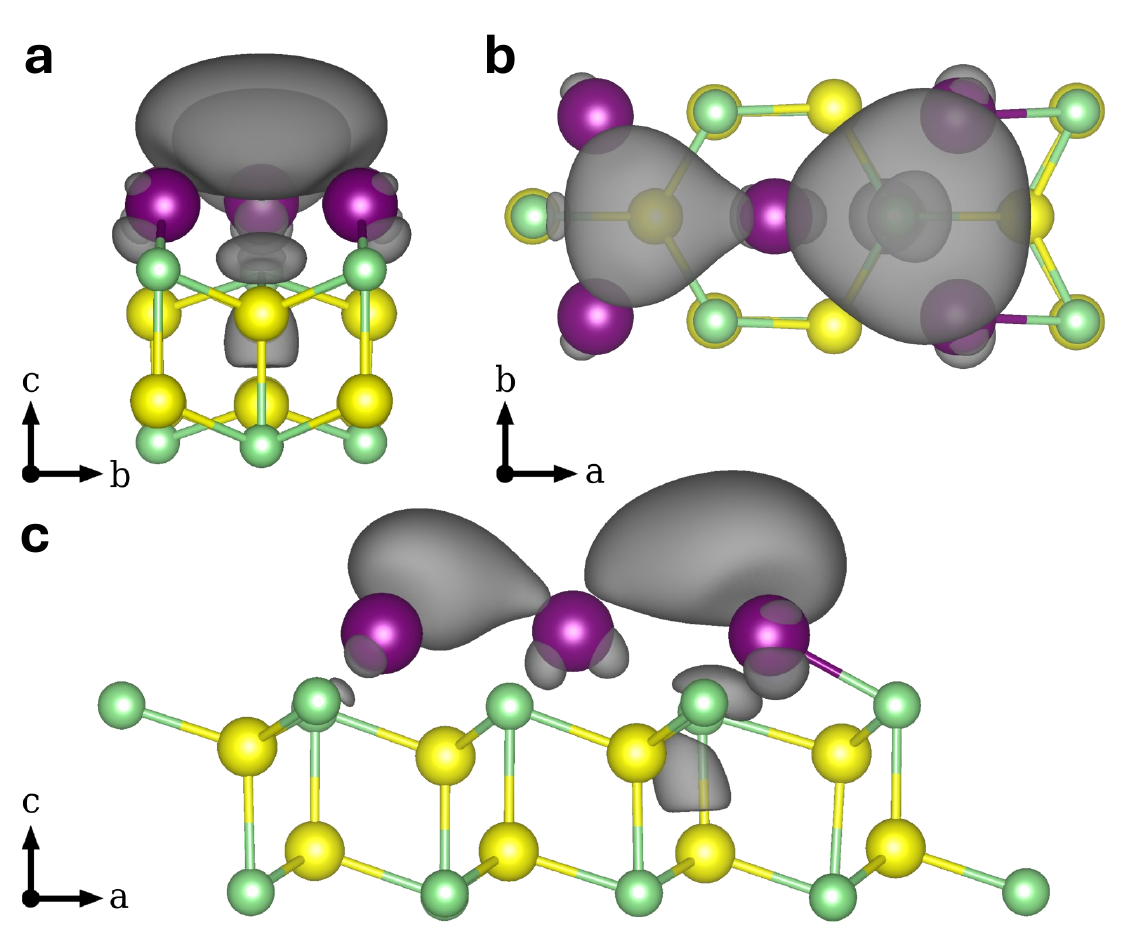}
	\end{center}
	\caption{{\bf Charge density around a single impurity.} We show with violet, green and yellow spheres Eu, Cd and As atoms, respectively. In grey we show the charge density around a single Eu impurity, located in between the Eu rows. We provide results close to the Fermi level (integrated from $-0.1$ to $+0.1$~eV around $E_F$), with lateral ({\bf a}, {\bf c}) and top ({\bf b}) views of the charge density.}
	\label{ChargeDensity}
\end{figure}

\begin{figure}
	\includegraphics[width=1\columnwidth]{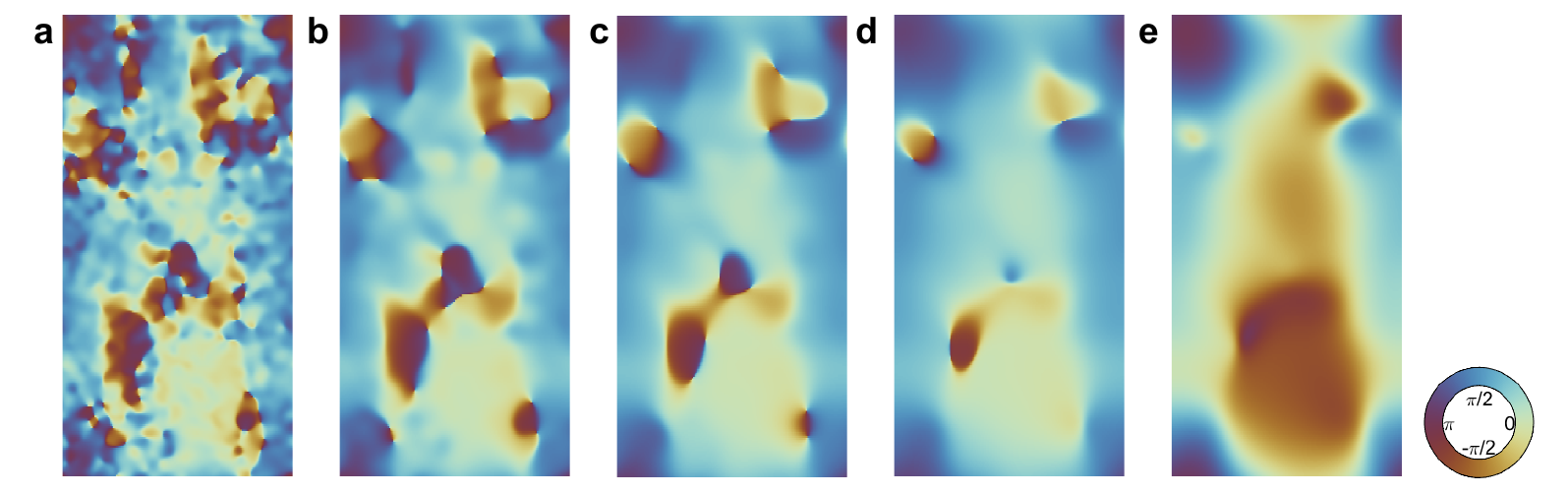}
	\caption{Phase maps obtained with different radiii in the Gaussian filter function (approximately the radius of the circle in Fig.\,\ref{SuplExampleFilter}{\bf c}). We take as an example the zero field map obtained at $+14$~mV. Radius goes from 1 ({\bf a}), 0.5  ({\bf b}), 0.33  ({\bf c}), 0.25  ({\bf d}) and 0.2  ({\bf e}) nm$^{-1}$. The phase follows the color pattern given at the bottom right of the figure. We take a radius of 0.33 ({\bf c}) in all figures showing the phase pattern.}
	\label{GaussianComp}
\end{figure}

In Fig.\,\ref{SuplModel2}{\bf d}, {\bf e},{\bf f} we present LDOS maps illustrating how the
phase dislocation positions rotate for different values of the anisotropic hopping
amplitude $t_p$.

\clearpage
\newpage

\clearpage
\newpage

\section*{Acknowledgments}
This work was supported by the Spanish Research State Agency (PID2023-150148OB-I00, TED2021-130546B-I00, PDC2021-121086-I00 and CEX2023-001316-M), by the Comunidad de Madrid through program MAG4TIC-CM (TEC-2024/TEC-380) and PhD thesis support PIPF-2023/TEC-30683 and PIPF-2023/TEC-30853 and by the European Research Council PNICTEYES grant agreement 679080 and VectorFieldImaging grant agreement 101069239. We acknowledge collaborations through EU program Cost CA21144 (Superqumap) and the EU through grant agreement No. 871106. We acknowledge segainvex for design and construction of STM and cryogenic equipment. We acknowledge the QUASURF project (ref. SI4/PJI/2024-00199) funded by the Comunidad de Madrid through the agreement to promote and encourage research and technology transfer at the Universidad Aut\'onoma de Madrid. Work at the Ames National Laboratory was supported by the U.S. Department of Energy (DOE), Office of Science, Basic Energy Sciences, Materials Sciences and Engineering Division. Ames National Laboratory is operated for the U.S.\,Department of Energy by Iowa State University under Contract No.\,DE-AC02-07CH11358. A.V. and R.V. acknowledge support by the Deutsche Forschungsgemeinschaft (DFG, German Research Foundation) for funding through TRR 288-422213477 (project A05, B05). R.P.M. acknowledge support from project ``Ampliación del uso de la mecánica cuántica desde el punto de vista experimental y su relación con la teoría, generando desarrollos en tecnologías cuánticas útiles para metrología y computación cuántica a nivel nacional'', BPIN 2022000100133 from SGR of MINCIENCIAS; Gobierno de Colombia, and W.J.H. acknowledges support  of Universidad Nacional de Colombia DIEB project ''Fortalecimiento Centro de Excelencia Tecnologías Cuánticas y sus Aplicaciones a Metrología'', Hermes code 67045.

\paragraph*{Author contributions.} \
\noindent  R.P.M., W.J.H. and A.L.Y. performed the microscopic calculations and proposed the analytical model relating phase slips to the DOS. A.V. and R.V. carried out the density functional theory (DFT) slab calculations and analyzed the symmetry and surface electronic properties of \(\text{EuCd}_2\text{As}_2\). R.S.B. performed the low-temperature scanning tunneling microscopy experiments under the supervision of I.G., E.H., and H.S.. P.G.T. calculated the phase slip trajectories and executed the quantitative fits to the analytical model, supervised by I.G., E.H., and H.S.. B.K., S.L.B., and P.C.C. conceived the material's investigation, synthesized the high-purity single crystals, and performed bulk characterization. Funding acquisition was coordinated by R.V., P.C.C., W.J.H., A.L.Y., I.G., and H.S. The manuscript was written by H.S., R.S.B., I.G., A.V., R.P.M., W.J.H., and A.L.Y., with critical revisions and contributions from all authors.

\paragraph*{Data availability.}
All data will be available on {OSF} {doi.org/10.17605/OSF.IO/EKYVQ} upon  publication.

\clearpage
\newpage

\setcounter{figure}{0}

\renewcommand{\thefigure}{S\arabic{figure} }

\section*{Supplementary Information}

\section{T\lowercase{unneling conductance vs bias voltage and interstitial impurities}}

\begin{figure}
	\includegraphics[width=0.8\textwidth]{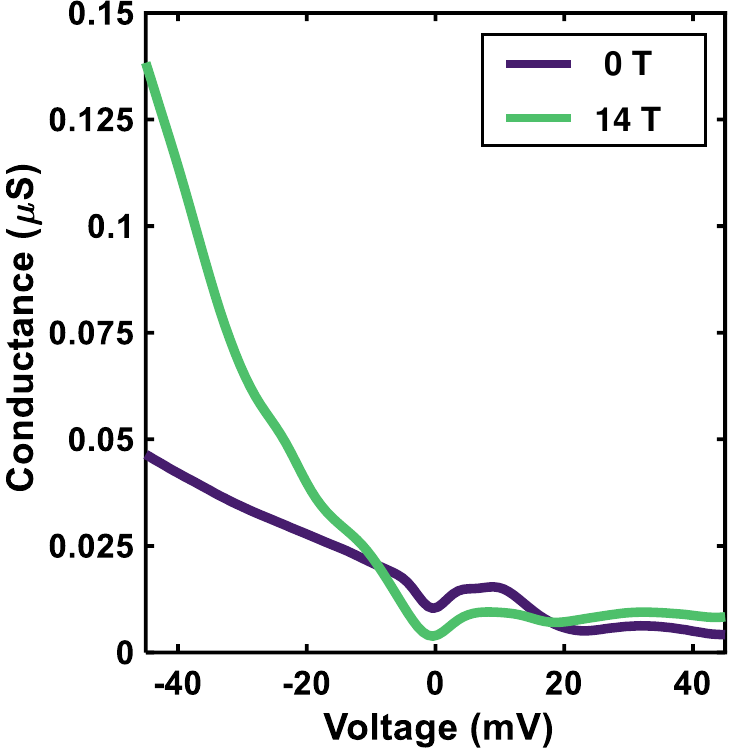}
	\caption{We show as violet and green colored lines the tunneling conductance as a function of the bias voltage taken at 0~T and 14~T, respectively. Note the strong increase in the tunneling conductance for negative bias, which corresponds to filled states in the electronic band structure. Note also the very low density of states for positive bias voltages and the presence of wiggles and peaks in the curves.}
	\label{SuplConductance}
\end{figure}

In Fig.\,\ref{SuplConductance} we show the tunneling conductance curves typically measured on the $2\times1$ surface at zero field and with a magnetic field of 14~T applied perpendicular to the surface of the sample. The tunneling conductance is much higher for negative bias voltages. This corresponds to filled states in the density of states and agrees with the shape of the electronic band structure in the valence bands close to the Fermi level (Fig.\,2{\bf a} of the main text). When applying a strong magnetic field, we find considerable increase in the tunneling conductance below the Fermi level (Fig.\,\ref{SuplConductance}, green curve). This suggests that bands below the Fermi level split and, as a consequence, new bands enter the bias voltage window.

\begin{figure}
	\includegraphics[width=1\textwidth]{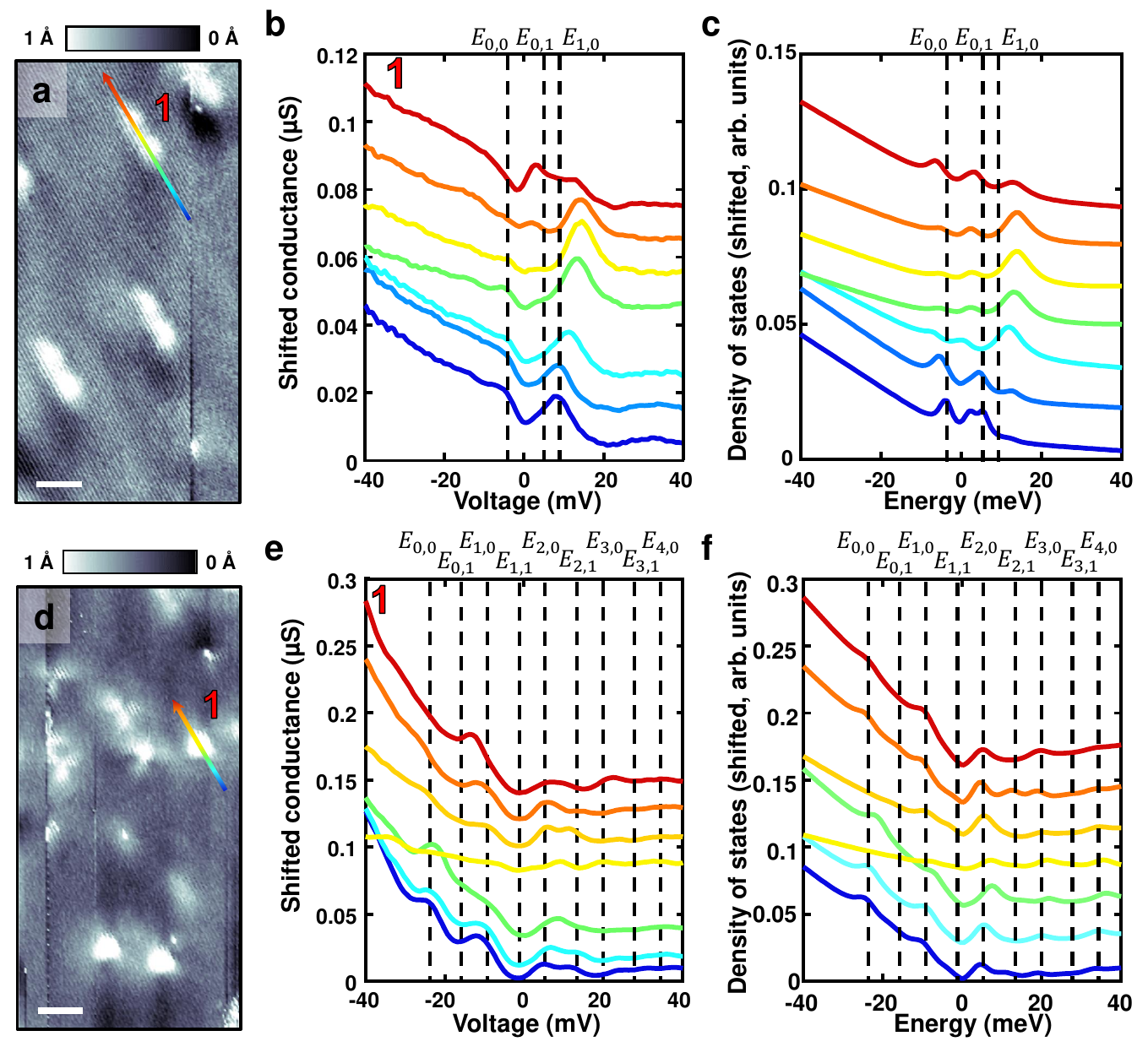}
	\caption{{\bf a} Topographic STM image at zero field (taken at $+$100~mV and 0.2~nA). Horizontal scale line is 10~nm long. The colored arrow marked as 1 shows the position where we took tunneling conductance curves shown in {\bf b}. The color scale of the arrow provides the positions at which we took the curves in {\bf b}. In {\bf b} we show the tunneling conductance at these positions. {\bf c} Calculations of the tunneling conductance vs energy, described in the text. Vertical dashed arrows provide the positions of discrete energy levels ($E_{l}$ and $E_{l,n}$ discussed in the text). Curves in {\bf b}, {\bf c} are vertically shifted for clarity. In {\bf d} we show a topography at 14~T on a different field of view and in {\bf e} the tunneling conductance curves taken along the arrow marked by 1 in {\bf d}. Further tunneling conductance maps are shown in Figs.\,\ref{SuplOtherStates},\ref{SuplOtherStatesScans},\ref{FullMap0},\ref{FullMap14}.}
	\label{Fig_S2States}
\end{figure}

Furthermore, we observe peaks in the tunneling conductance close to the Fermi level at zero as well as under magnetic fields. These peaks strongly evolve with position, following the topography. Let us take, for example, the area close to the arrow marked as ``1'' in Fig.\,\ref{Fig_S2States}{\bf a}. Roughly at the center of the arrow there is one Eu interstitial. We observe a large charge puddle in the topography. The topography is the integral of the tunneling conductance between the Fermi level and the applied bias voltage (here $+$100~mV). Therefore, it integrates the density of states in energy at a fixed position over all features observed between zero bias and the bias voltage. Inside the charge puddle shown as a white blob around ``1'' in Fig.\,\ref{Fig_S2States}{\bf a} there are numerous peaks in the bias voltage dependent tunneling conductance, whose height and bias voltage position changes as a function of position.

The tunneling conductance vs bias voltage curves along the arrow are shown in Fig.\,\ref{Fig_S2States}{\bf b}. Colors correspond to the colors marked along the arrow in Fig.\,\ref{Fig_S2States}{\bf a}. We see that there are at least three peaks which appear on different positions. A similar result, with peaks lying at slightly different positions, is found in all impurities of the image. We show further examples in Fig.\,\ref{SuplOtherStates}{\bf a},{\bf b}, {\bf c}. Under magnetic fields (shown in Fig.\,\ref{Fig_S2States}{\bf d}, {\bf e} and in Fig.\,\ref{SuplOtherStates}{\bf d},{\bf e}, {\bf f}, {\bf g},  {\bf h}  for another field of view), the situation is very similar. There are sets of peaks located at certain bias voltages, which change as a function of the position.

States around isolated impurities have been studied using STM in several semiconducting systems\,\cite{PhysRevLett.71.1176,PhysRevLett.72.840,PhysRevB.66.161306}. Our observations coincide with those found previously. We remark that most data are taken in large gap semiconductors, and the peaks develop mostly at bias voltages of several hundreds of meV, whereas here we mostly find peaks below 100 meV.

These peaks are clearly due to LDOS variations of impurity states which are partly confined close to the impurity site.

To better understand these peaks we consider a two-dimensional electron gas with potential wells that separate different locations along one direction, specifically, the direction of the Eu rows. We assume the existence of three regions, denoted by $j=L, c$, and $R$. These regions exhibit distinct electronic properties and are separated by small barriers, denoted as $t_a$ and $t_b$, located at $y=a$ and $y=b$, respectively. Along the $x$ direction (which is perpendicular to the Eu rows), we impose confinement using a harmonic potential given by $V(x)=\omega_0 x^2 /2$. To determine the differential conductance, we model the Green function for each region $j$ as

\[
g\left( x,x^{\prime },y,y^{\prime }\right) =\sum\limits_{n}\phi _{n}^{\ast
}\left( x^{\prime }\right) \phi _{n}\left( x\right) g_{n}\left( y,y^{\prime
}\right) ,
\]

where $n$ represents the level of the quantum harmonic oscillator, with the associated wave function $\phi _{n}\left( x\right) $. The Green function $g_{n}\left( y,y^{\prime }\right)$  solves the inhomogeneous Schr\"{o}dinger equation

\[
\left( H-\hbar \omega _{0}\left( n+\frac{1}{2}\right) -E-E_{F}\right)
g_{j,n}\left( y,y^{\prime }\right) =\delta \left( y-y^{\prime }\right) ,
\]%
$g_{j,n}\left( y,y^{\prime }\right) $ can be calculated as

\[
g_{n}\left( y,y^{\prime }\right) =A\varphi _{jn}^{<}\left( y_{<}\right)
\varphi _{jn}^{>}\left( y_{>}\right) ,
\]%
where $\varphi _{jn}^{<\left( >\right) }\left( y_{<\left( >\right) }\right) $
is the solution for the  potential well $j,$ here $\varphi _{>(<),n}\left(
y\right) $ satisfies the  Neumann boundary conditions at the right (left)
edge of the well $j$. For the lowest quantum levels $m$, we can express the
following Green function for each region of the nanowire 
\[
g_{j,n}\left( y,y^{\prime }\right) \cong \sum\limits_{m=0}^{m_{j}}\frac{%
\varphi _{j,n}^{\ast }\left( y^{\prime }\right) \varphi _{j,n}\left(
y\right) }{E-\epsilon _{j,nm}+i\Gamma _{j,nm}},\;j=L,c,R,
\]%
where $m_{j}=1$, for $j=L,R$, and 0 for $j=c,$ Here $\epsilon _{j,nm}$\ are
the energies of nanowire for the section $j$%
\begin{equation}
\epsilon _{j,nm}=\left( n+\frac{1}{2}\right) \hbar \omega _{0}+\frac{\hbar
^{2}}{2m^{\ast }}\left( \frac{m\pi }{W}\right) ^{2}-E_{j,F}.
\end{equation}
where $W$ is the width of $j$ region. The regions $L$ and $R$ are coupled with
central region with $t_{a}$ and $t_{b}$ hoppings respectively.  From Dyson
equation we can find the perturbative Green function of each region as 
\[
G_{i,n}\left( y,y^{\prime }\right) =g_{i,n}\left( y,y^{\prime }\right)
+g_{i,n}\left( y,y_{0}\right) \Sigma _{ij}G_{j,n}\left( y_{0},y^{\prime
}\right) \text{, }
\]%
$i,j=L,c,R.$ For $y_{0}=a\left( b\right) $ the self energies are $\Sigma
_{Lc}=t_{a}\;\left( \Sigma _{cR}=t_{b}\right) .\;$\ Average in $x$
coordinate, $G_{j}\left( y,y^{\prime }\right) =\left\langle G_{j}\left(
x,x,y,y^{\prime }\right) \right\rangle $, the total Green function is  
\begin{equation}
G_{j}\left( y,y^{\prime }\right) =\sum\limits_{n=0}^{n_{\text{max}%
}}G_{j,n}\left( y,y^{\prime }\right) .
\end{equation}%
When $B=0$, $n_{max}=0,$ thus, we have two levels in $L$\ and $R$ region,
and one level in $c$ region. When  $\mathbf{B}=B\hat{k}$, and with the gauge
for the vectorial potential $\mathbf{A}=-By\hat{\imath},$ the spectrum of
energy is modified, where Landau levels appear with frequency  $
\omega=\sqrt{\omega _{0}^{2}+\omega _{c}^{2}}$. Therefore the energy $%
\epsilon _{j,nm}$ is replaced by 

\begin{equation}  
E_{n,m}=\left( n+\frac{1}{2}\right) \hbar \omega+\frac{\hbar ^{2}}{%
2m_{r}}\left( \frac{m\pi }{W}\right) ^{2}-E_{j,FB},
\end{equation}
with  $\omega _{c}$ the cyclotron frequency. We have various
energy levels due to magnetic confinement. Therefore, in this case we have $%
m_{L,R}=0,1,m_{c}=0,$ and $n=0,1,..,n_{\text{max}}$, where we use a cutoff
of $n_{\text{max}}=4$.     

The differential conductance in the region $j$ is calculated as

\begin{equation}
\sigma _{j}(V,y)=\sigma _{0}T_{j,y}\left( eV\right) +\sigma_B,
\end{equation}%
with

\begin{equation}
T_{j,y}\left( eE\right) =-\frac{1}{\pi }\mathtt{Im}\left(\sum\limits_{n=0}^{n_{\text{max}}}t_{j,n}^{2}G_{j,n}\left( y,y \right)\right) ,
\end{equation}
$t_{j,n}$ the hopping between the tip and the channel $n$ of the nanowire. Here we have included the conductance $\sigma_{B}$ to model the background.

For example, in Fig.\,\ref{Fig_S2States}{\bf c}, {\bf f} we show the calculated tunneling conductance curves at different defects. In the area labeled as ``1'' in Fig.\,\ref{Fig_S2States} we find a large peak at about $+13$~mV at the center of defects and two further peaks at smaller bias voltages, $+4$ and $-5$~mV.

Another view of the same states can be obtained using a color scale for the conductance, as shown in Fig.\,\ref{SuplOtherStatesScans}{\bf a,b}. We see here clearly the evolution of the levels with distance.

\section{ L\lowercase{ocalized states at zero field and density of states patterns under magnetic fields}}

\begin{figure}
	\includegraphics[width=1\textwidth]{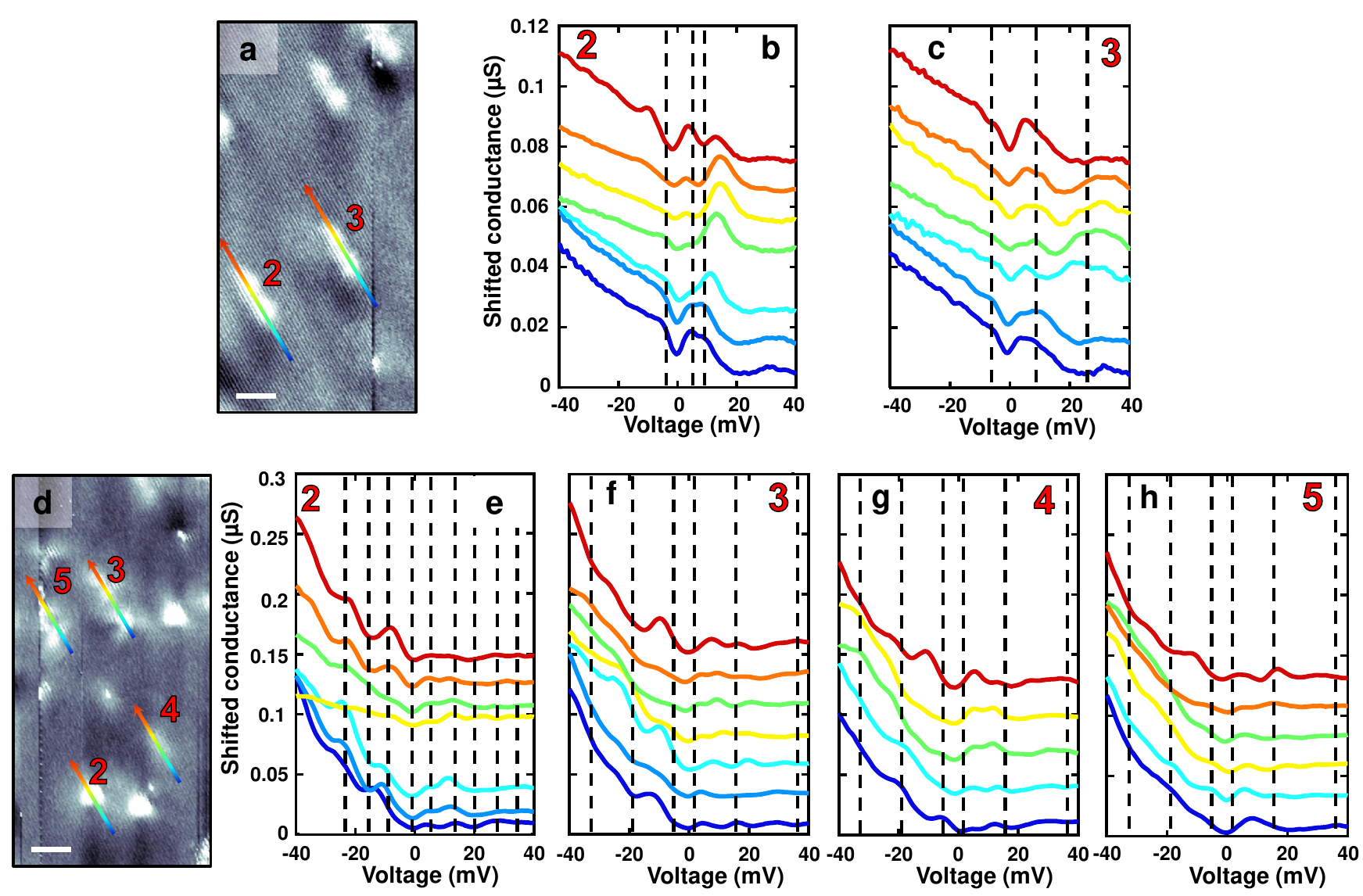}
	\caption{(a) Topographic STM image taken at 100~mV and 0.2~nA. Horizontal white bar is 10~nm long. In (b,c) we show as colored lines tunneling conductance curves taken along the defect marked as 2 and 3 in (a). The same plots are made for data taken at 14~T in (d-h). Curves are shifted vertically for clarity in (b,c,e-h). Dashed black lines indicate energy levels as discussed in the text.}
	\label{SuplOtherStates}
\end{figure}

\begin{figure}
	\includegraphics[width=1\columnwidth]{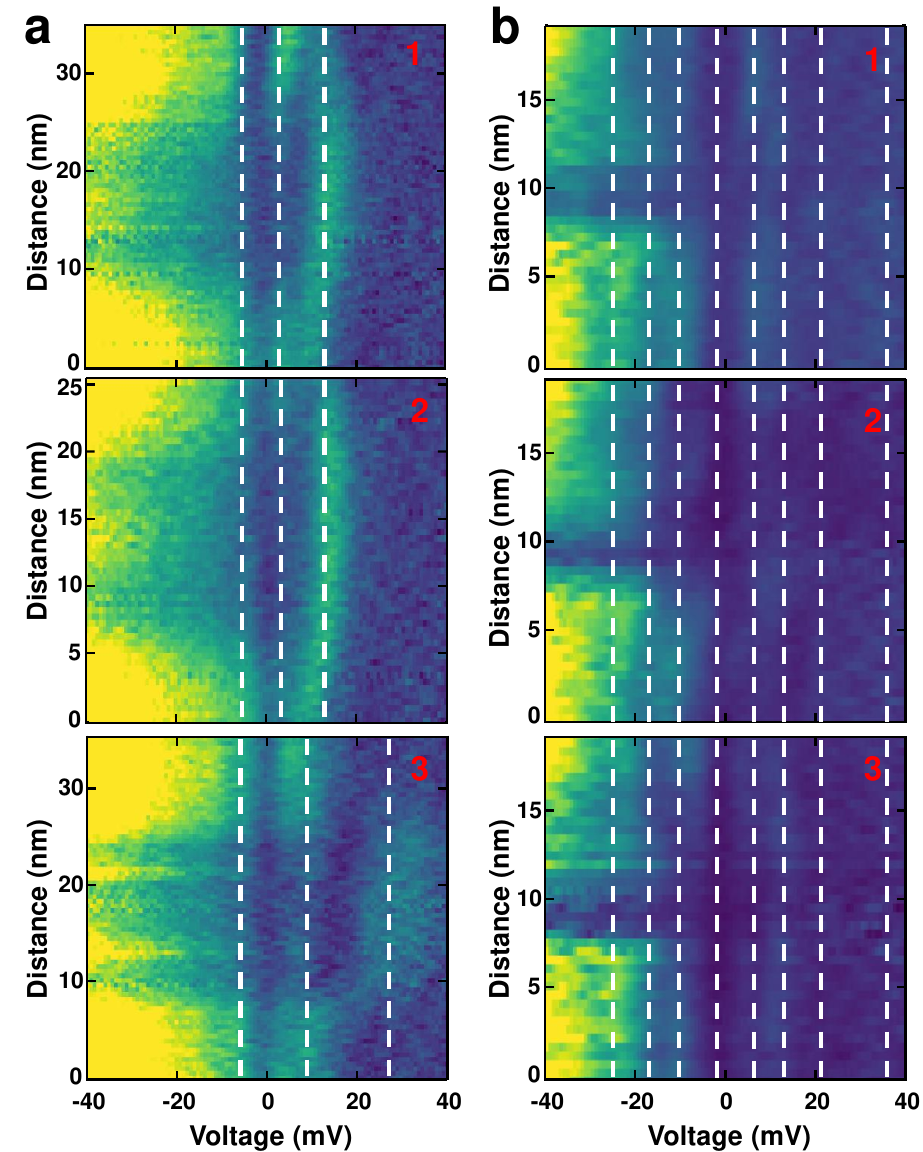}
	\caption{Conductance is shown as a color scale, as a function of the distance (y-axis) and bias voltage (x-axis) for zero magnetic field (a) and for a field of 14 T (b). Dashed white lines are the energy levels discussed in the text.}
	\label{SuplOtherStatesScans}
\end{figure}

We present tunneling conductance maps as a function of the bias voltage at zero field and at 14~T in Figs.\,\ref{FullMap0},\,\ref{FullMap14}. Let us start by discussing the dependence on the bias voltage of the charge puddle marked with ``3'' at zero field in Fig.\,\ref{SuplOtherStates}{\bf a}, which corresponds to the red dashed line in the central part of the panels shown in Fig.\,\ref{FullMap0} (see white arrow). We distinguish three maxima at negative bias voltages which turn into two maxima at about $+$5~mV and one broad level at about $+$25~mV. Interestingly the three maxima remain well defined at negative bias voltages. At negative bias voltages, the tunneling conductance is large, suggesting that there are electronic states everywhere on the surface. This shows that there is a strong variation of the chemical potential around an impurity, which moves towards negative bias in the surrounding of this impurity. The rest of the impurities, as for instance the one marked by ``2'' in Fig.\,\ref{SuplOtherStates}{\bf a} present similar variations.

Under magnetic fields, these variations in the chemical potential produce similar patterns. States are more intricate and distribute on a larger scale of energies. Previous work shows that when a localized state is close to a potential minimum, its radius increases with the bias voltage, and vice versa when it is close to a potential maximum in a two-dimensional electron gas\,\cite{PhysRevLett.109.166407}. We identify several positions in the tunneling conductance maps (Fig.\,\ref{FullMap14}) with elongated or semi-circular patterns which extend or decrease their size with the bias voltage. Some are highlighted with white arrows in Fig.\,\ref{FullMap14}.

\section{B\lowercase{and representation analysis of monolayer} {E\lowercase{u}C\lowercase{d}$_2$A\lowercase{s}$_2$}}

We have analyzed the possible orbital types at several Wyckoff positions in the monolayer including a Eu interstitial impurity of EuCd$_2$As$_2$. To increase the distance between impurities we placed them in opposite corners of the enlarged $2\times2\times1$ monoclinic unit cell where every second row of Eu was removed. This setup results in the space group $P2_1/m$ (No. 11). We used the setup and k-points of this space group, consistent with the Bilbao crystallographic server tables\,\cite{Aroyoxo5013,AroyoPerezMato}.

If we assume non-magnetic Eu, the resulting Wyckoff positions for Eu are then $2e$ and twice $4f$, which yield all 10 Eu atoms of the surface, where the former describes the impurity Eu atoms. From symmetry perspective only the $2e$ positions form the maximal subgroup $m$. The character table for $m$ shows the orbital types in Tab. \ref{tab:EuImpurity}.

By using the representations for the orbitals at the Fermi level, we identify degeneracies given by group theory and allowed accidental crossings. To obtain non-trivial topology at the Fermi level, there must be a mixture of even and odd parity bands. By investigating this in further detail we obtain the band representations of A' and A''. This is done by using the \verb|BANDREP| tool of the bilbao crystallographic server, for space group $P2_1/m$ (No. 11) and \verb|Wyckoff TR| position 2e - the \verb|TR| stands for time reversal, which is valid in this case as the $4f$ momenta are excluded of the calculation. 
The result is shown in Tab.\,\ref{tab:EuImpurity:bandrep}.
\begin{table}[!h]
\centering
\footnotesize
\begin{tabular}{c|c|c|c}
\hline
$m$   & E & $\sigma$ & orbital type                               \\ \hline\hline
A'  & 1 & 1        & $s, p_x, p_y, d_{x^2-y^2}, d_{xy}, d_{z^2}$     \\ \hline
A'' & 1 & -1       & $p_z, d_{xz}, d_{yz}$                      \\ \hline
\end{tabular}
\caption{Possible orbital representations for the monolayer with impurity exactly at the impurity, i. e. for Wyckoff position 2e with site symmetry $m$.}
\label{tab:EuImpurity}
\end{table}

\begin{table*}[!h]
\centering
\scriptsize
\begin{tabular}{|c|c|c|c|c|c|c|c|c|c|c|c|}
\hline
 $P2_{1}/m$          & $\Gamma$                                                                              & F         & B                                                                             & V                                                                   & D         & GP         & $\Gamma$                                                                              & $\Lambda$                                                                       & Z        & G             & D         \\ \hline
\textbf{k} & (0,0,0)                                                                               & (u,0,w)   & (0,0,.5)                                                                      & (0,v,.5)                                                            & (0,.5,.5) & (u,v,w)    & (0,0,0)                                                                               & (0,v,0)                                                                         & (0,.5,0) & (u,.5,w)      & (0,.5,.5) \\ \hline
A'(1)      & \begin{tabular}[c]{@{}c@{}}$\Gamma_1^{+}(1)$\\ $\oplus$$\Gamma_2^{-}(1)$\end{tabular} & 2F$_1$(1) & \begin{tabular}[c]{@{}c@{}}B$_1^{+}(1)$\\ $\oplus$B$ _2^{-}(1)$\end{tabular}  & \begin{tabular}[c]{@{}c@{}}V$_1$(1)\\ $\oplus$V$_2$(1)\end{tabular} & D$_1$(2)  & 2GP$_1$(1) & \begin{tabular}[c]{@{}c@{}}$\Gamma_1^{+}(1)$\\ $\oplus$$\Gamma_2^{-}(1)$\end{tabular} & \begin{tabular}[c]{@{}c@{}}$\Lambda_1$(1)\\ $\oplus$$\Lambda_2$(1)\end{tabular} & Z$_1$(2) & G$_1$G$_2$(2) & D$_1$(2)  \\ \hline
A''(1)     & \begin{tabular}[c]{@{}c@{}}$\Gamma_1^{-}(1)$\\ $\oplus$$\Gamma_2^{+}(1)$\end{tabular} & 2F$_2$(1) & \begin{tabular}[c]{@{}c@{}}B$_1^{-}(1)$\\ $\oplus $B$ _2^{+}(1)$\end{tabular} & \begin{tabular}[c]{@{}c@{}}V$_1$(1)\\ $\oplus$V$_2$(1)\end{tabular} & D$_1$(2)  & 2GP$_1$(1) & \begin{tabular}[c]{@{}c@{}}$\Gamma_1^{-}(1)$\\ $\oplus$$\Gamma_2^{+}(1)$\end{tabular} & \begin{tabular}[c]{@{}c@{}}$\Lambda_1$(1)\\ $\oplus$$\Lambda_2$(1)\end{tabular} & Z$_1$(2) & G$_1$G$_2$(2) & D$_1$(2)  \\ \hline
\end{tabular}
    \caption{Band representations for Wyckoff positions 2e. To account for spin one has to multiply every dimension, i. e. the number in brackets, with a $\times 2$.}
    \label{tab:EuImpurity:bandrep}
\end{table*}

\begin{table*}[!h]
\centering
\tiny
\begin{tabular}{|c|c|c|c|c|c|c|c|c|c|c|c|}
\hline
  $P2_{1}/m1'$                & $\Gamma$                                                                                                             & F                        & B                                                                                                & V                        & D                       & GP                         & $\Gamma$                                                                                                             & $\Lambda$                            & Z                       & G                                                                                                  & D                       \\ \hline
\textbf{k}                                                                                  & (0,0,0)                                                                                                              & (u,0,w)                  & (0,0,.5)                                                                                         & (0,v,.5)                 & (0,.5,.5)               & (u,v,w)                    & (0,0,0)                                                                                                              & (0,v,0)                              & (0,.5,0)                & (u,.5,w)                                                                                           & (0,.5,.5)               \\ \hline
\begin{tabular}[c]{@{}c@{}}${}^1\!\bar{E}{}^2\!\bar{E}(4)$\\ \end{tabular} & \begin{tabular}[c]{@{}c@{}}$\bar{\Gamma}_3\bar{\Gamma}_4(2)$\\ $\oplus \bar{\Gamma}_5\bar{\Gamma}_6(2)$\end{tabular} & $2\bar{F}_3\bar{F}_4(2)$ & \begin{tabular}[c]{@{}c@{}}$\bar{B}_3\bar{B}_4(2)$\\ $\oplus \bar{B}_5\bar{B}_6(2)$\end{tabular} & $2\bar{V}_3\bar{V}_4(2)$ & $\bar{D}_2\bar{D}_2(4)$ & $2\bar{GP}_2\bar{GP}_2(2)$ & \begin{tabular}[c]{@{}c@{}}$\bar{\Gamma}_3\bar{\Gamma}_4(2)$\\ $\oplus \bar{\Gamma}_5\bar{\Gamma}_6(2)$\end{tabular} & $2\bar{\Lambda}_3\bar{\Lambda}_4(2)$ & $\bar{Z}_2\bar{Z}_2(4)$ & \begin{tabular}[c]{@{}c@{}}$\bar{G}_3\bar{G}_3(2)$ \\ $\oplus \bar{G}_4\bar{G}_4(2)$\end{tabular} & $\bar{D}_2\bar{D}_2(4)$ \\ \hline
\end{tabular}
\caption{The effect of spin orbit coupling on the Wyckoff position $2e$ for the band representations without magnetism.}
\label{tab:EuImpurity:spinorRep}
\end{table*}

From Tab.\,\ref{tab:EuImpurity:bandrep} we obtain that the degeneracies of the high symmetry points in the Brillouin zone are located especially at D and Z. The only high symmetry line (plane) with degeneracies is G.

Next step is then to introduce a spinor representation $\rho_{sp}$ in the site symmetry group $m$.
This only requires knowledge about $m$, as $m$ (with time reversal symmetry) has here only one representation, see Tab. \ref{tab:EuImpurity:spinorRep}.

From the tables and the bandstructure (Fig.\,\ref{fig:EuImpurity:bandStructure}), we identify a fourfold degenerate point at Z and at D.  Interestingly, B is not degenerate, there is a small shift by $1$meV. The high symmetry plane G, i. e. $(u,.5,w)$, has without SOC a fourfold degeneracy (including the spin).
Upon applying SOC this planar degeneracy gets lifted to two twofold degeneracies.
All in all, we remain with two symmetry enforced fourfold crossings, one at $Z$ and one at $D$. We find no non-trivial topological indices, only accidental crossings. If we assume in-plane Eu alignment of spins, following the c-axis antiferromagnetic order of the bulk, we find band structures which are very similar to the one discussed above (Fig.\,\ref{fig:EuImpurity:bandStructure}). We could not find any non-accidental crossings either.

\section{M\lowercase{ultifractal properties of the electronic density of states}}

The interplay between localization and interaction around defects in semiconductors provides an electronic wavefunction pattern that is not random, but multifractal\,\cite{RevModPhys.80.1355,PhysRevLett.67.607}.  Multifractality has been analyzed in electronic systems at surfaces, in doped semiconductors and in two-dimensional correlated electron systems\,\cite{doi:10.1126/science.1183640,PhysRevB.108.104205,MIRLIN2000259,PhysRevB.83.184206,CARNIO2019141,MORGENSTERN20121795,PhysRevLett.97.046803}. The spectrum of multifractal exponents is given by the function $f(\alpha)$, which is obtained from the tunneling conductance maps by calculating locations with the same exponent $\alpha$ in the dependence of the square of the wavefunction with size\,\cite{pepe2020, postolova2020,morgenstern2003, terao1996,doi:10.1126/science.1183640}. For an entirely random two-dimensional image, $f(\alpha)$ is a point $f(\alpha_{max})=2$ at $\alpha_{max}$ = 2. $f(\alpha)$ is obtained by dividing a LDOS map into boxes and calculating the LDOS weight in each box. The variation of the weight with the size of the box follows a power law with exponent $\alpha$. When the LDOS is randomly distributed on a two-dimensional map, the exponent is just 2. When the LDOS distribution shows patches and voids, $\alpha$ deviates from 2. The function $f(\alpha)$ describes this deviation. It is an inverted parabola with maximum at $\alpha_{max}$, deviating from 2 when LDOS presents non-random patterns. Multifractality is a consequence of the moments of the wavefunction presenting a power law scaling with the size of the system and leads to a divergent correlation length. In areas where the LDOS $\rho(\mathbf{r})$ scales with size as $L^{-\alpha}$, the average of the LDOS (the inverse participation ratio) scales as $L^{f(\alpha)}$.

To calculate the multifractal properties we use the box counting method described in Ref.\,\cite{pepe2020,chhabra1989}.

\begin{figure*}
	\includegraphics[width=0.82\textwidth]{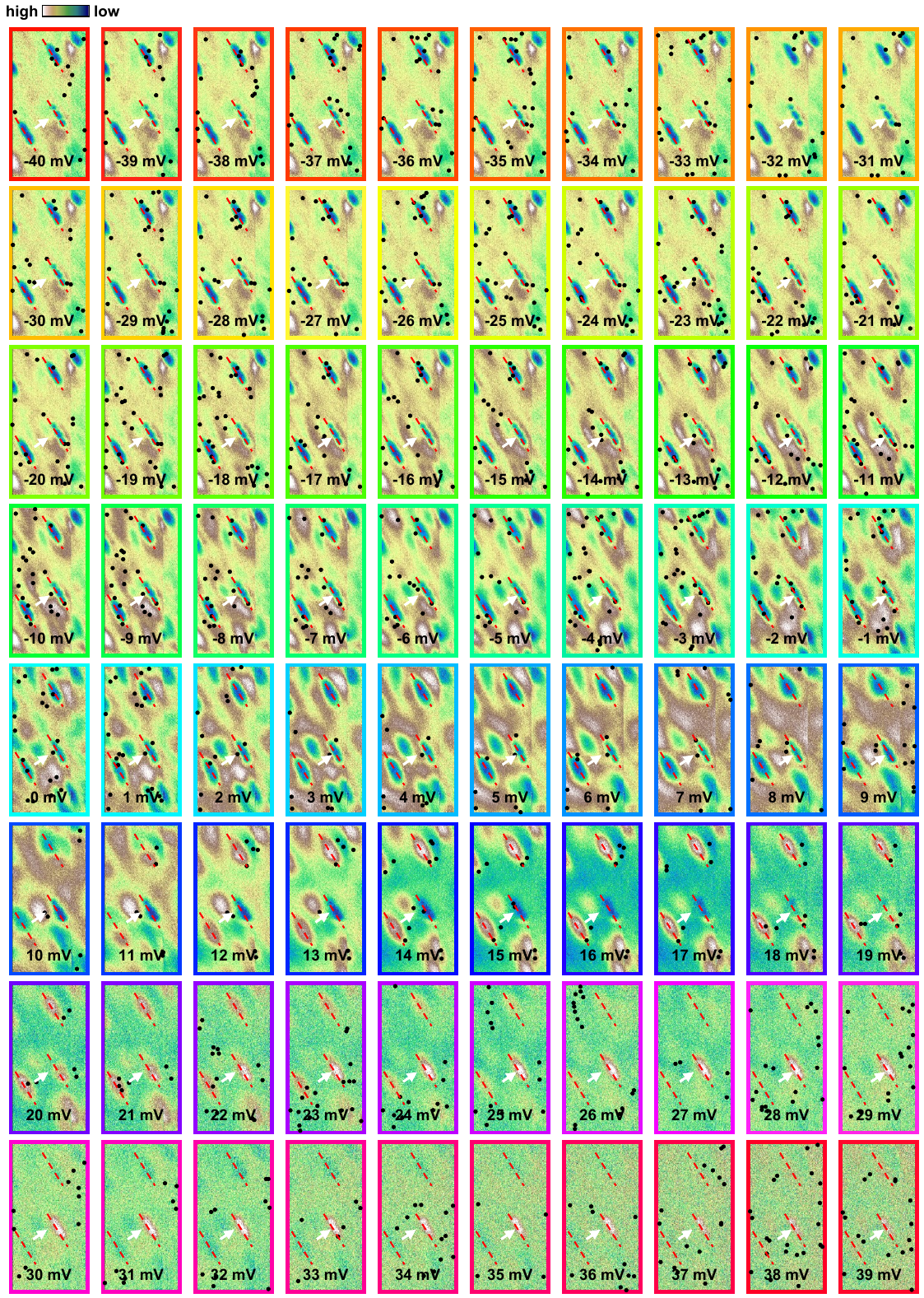}
	\caption{Set of tunneling conductance maps as a function of the bias voltage at zero magnetic field. The full sequence is available as a Supplementary Video. The color scale is adjusted at each bias voltage to obtain maximal contrast. Each panel is framed with the color also used in Fig.\,\ref{FigureS7}. Black dots provide positions for phase slips, discussed below. White lines highlight a pattern discussed in the text. Dashed red lines provide the lines shown in Figs.\,\ref{Fig_S2States} and \ref{SuplOtherStates}.}
	\label{FullMap0}
\end{figure*}

\begin{figure*}
	\includegraphics[width=0.82\textwidth]{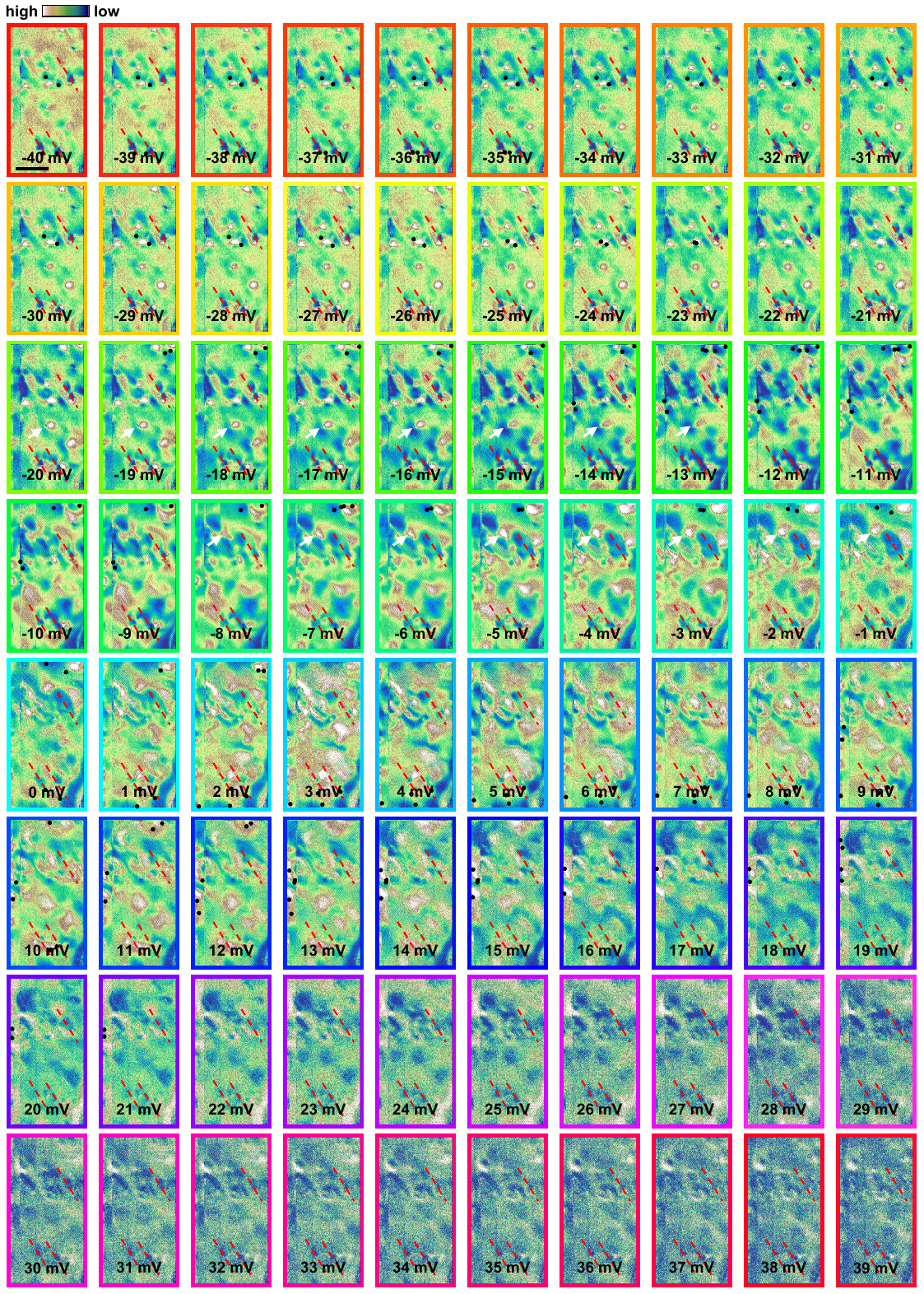}
	\caption{Set of tunneling conductance maps as a function of the bias voltage at 14\,T, in another field of view as compared to the zero field data. The full sequence is available as a Supplementary Video. The color scale is adjusted at each bias voltage to obtain maximal contrast. Each panel is framed with the color also used in Fig.\,\ref{FigureS7}. Black dots provide positions for phase slips, discussed below. White lines highlight a pattern discussed in the text. Dashed red lines provide the lines shown in Figs.\,\ref{Fig_S2States} and \ref{SuplOtherStates}.}
	\label{FullMap14}
\end{figure*}

\begin{figure*}[h!]
    \centering
    \includegraphics[width=\linewidth]{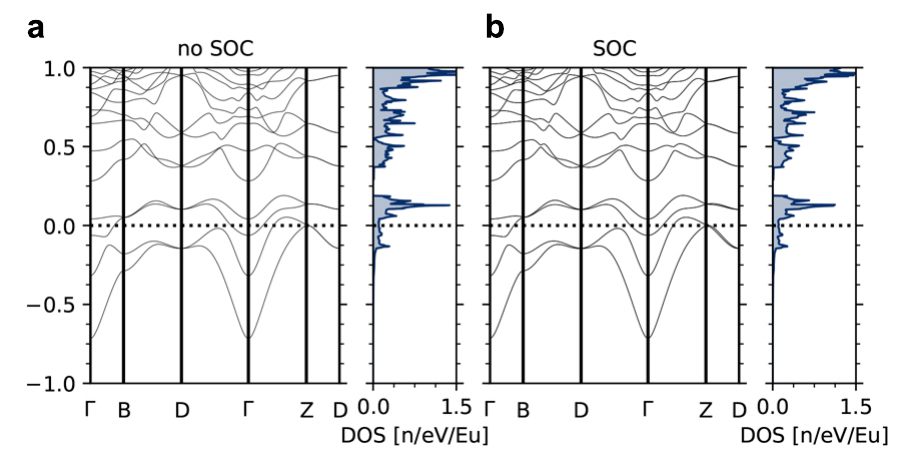}
    \caption{
    Band structure of EuCd$_2$As$_2$ with a Eu impurity without SOC ({\bf a}) and with SOC ({\bf b}). 
    Note for example a gap opening between $\Gamma$ and Z. Furthermore, a (fourfold) symmetry line between $Z$ and $D$ is lifted to two (twofold) lines.}
    \label{fig:EuImpurity:bandStructure}
\end{figure*}

\begin{figure}
	\includegraphics[width=1\columnwidth]{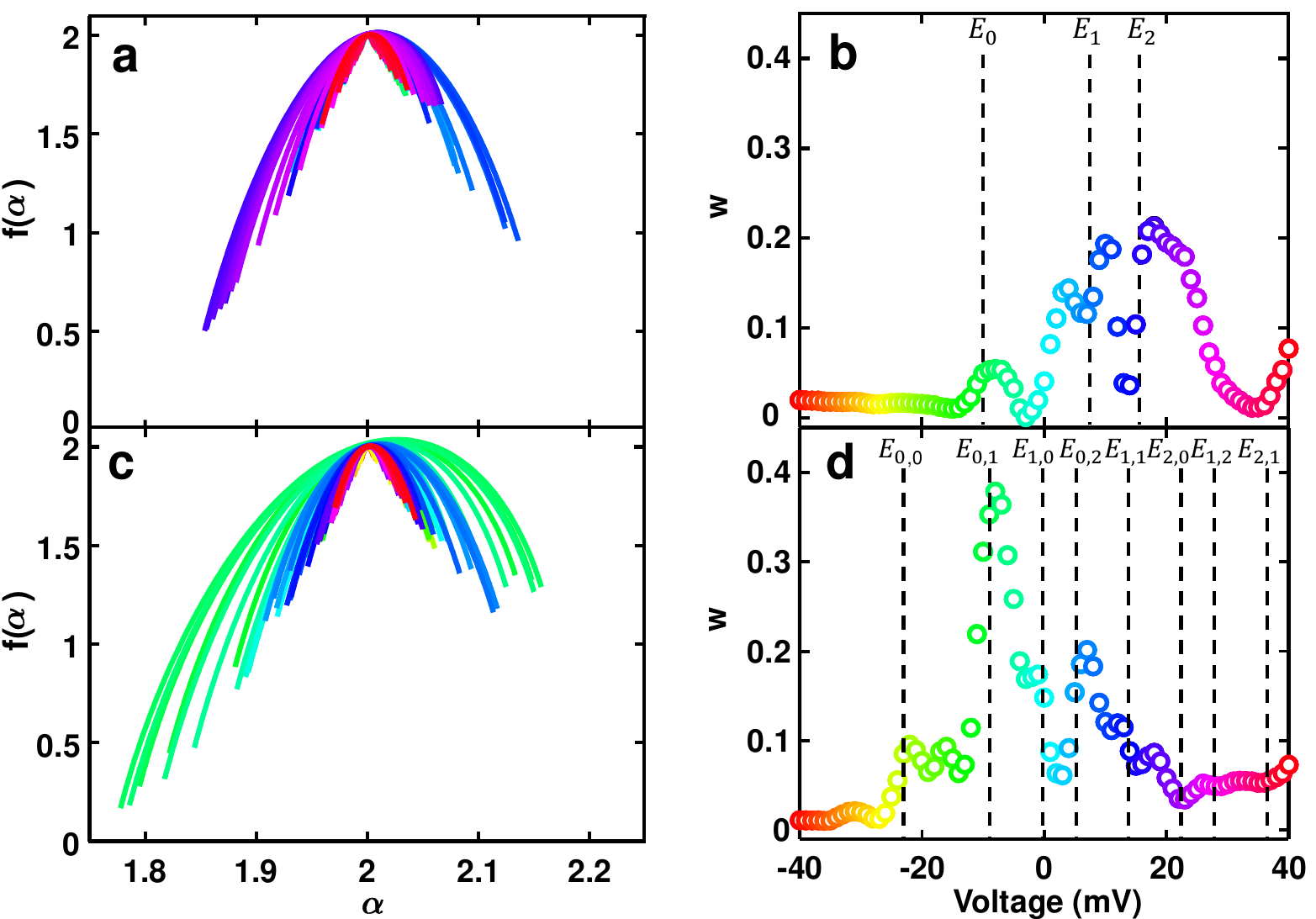}
	\caption{{\bf a} Function $f(\alpha)$ obtained from tunneling conductance maps at zero magnetic field. The color scale goes from red at -40~mV to dark red at +40~mV, as shown in {\bf b}. In {\bf b} we represent as colored dots the width of the parabolae shown in {\bf a} as a function of the bias voltage. Dashed lines provide the position of discrete states within the field of view. The same plots for a magnetic field of 14~T are shown in {\bf c},{\bf d}.}
	\label{FigureS7}
\end{figure}

In Fig.\,\ref{FigureS7}{\bf a} we show $f(\alpha)$ at zero magnetic field, obtained from the tunneling conductance images at different bias voltages (shown in Fig.\,\ref{FullMap0}). The distribution of values of $\alpha$, w,  as a function of the bias voltage is shown in Fig.\,\ref{FigureS7}{\bf b}. We find a randomly distributed LDOS far from the Fermi level (w $\approx 0$ and $f(\alpha)=2$). However, at the energy of each localized level, we find an increasing distribution of values of $\alpha$. At the same time, $f(\alpha)$ broadens from a single point with $f(\alpha_{max})=2$ into a parabola with a width w. The resulting w is particularly large with an applied magnetic field (Fig.\,\ref{FigureS7}{\bf c}, {\bf d}).

STM results in doped semiconductors and in two-dimensional electron gases show that wavefunctions present a multifractal exponent at localization\,\cite{doi:10.1126/science.1183640,PhysRevB.108.104205,morgenstern2003,MORGENSTERN20121795,PhysRevResearch.3.013022,Rubio-Verdu2020}. The deviation of the multifractal exponent $\alpha_{max}$ from the random value ($\alpha_{max}=2$) is often quite small, only w $\approx 0.01$ in two-dimensional quantum Hall electron gases \cite{morgenstern2003} or 0.1 for small concentration of impurities in semiconductors \cite{morgenstern2003}. Here we find a similar deviation, of 0.2 at zero field and up to 0.4 under magnetic fields.

\clearpage
\newpage


\clearpage
\newpage

\bibliographystyle{naturemag}
\bibliography{Biblio}

\end{document}